\documentclass[aps,showpacs]{revtex4}

\usepackage{amssymb}
\usepackage{graphicx}
\usepackage[T1]{fontenc}
\usepackage{euscript}

\begin{document}
\title{\bf
Non-minimal coupling for the gravitational
and electromagnetic fields:\\ black hole solutions and solitons}

\author{Alexander B. Balakin\footnote{Email: e-mail:
Alexander.Balakin@ksu.ru}}
\affiliation{Kazan State University, Kremlevskaya str.,
18, 420008, Kazan,  Russia,}
\author{Vladimir V. Bochkarev\footnote{Email: Vladimir.Bochkarev@ksu.ru}}
\affiliation{Kazan {} State University, Kremlevskaya str.,
18, 420008, Kazan,  Russia,}
\author{Jos\'e P. S. Lemos\footnote{Email: lemos@fisica.ist.utl.pt}}
\affiliation{Centro Multidisciplinar de {}Astrof\'{\i}sica - CENTRA \\
Departamento de F\'{\i}sica, Instituto Superior T\'ecnico - IST,\\
Universidade T\'ecnica de Lisboa - UTL,\\
Av. Rovisco Pais 1, 1049-001 Lisboa, Portugal.}

\begin{abstract}
Using a Lagrangian formalism, a three-parameter non-minimal
Einstein-Maxwell theory is established. The three parameters, $q_1$,
$q_2$ and $q_3$, characterize the cross-terms in the Lagrangian,
between the Maxwell field and terms linear in the Ricci scalar, Ricci
tensor, and Riemann tensor, respectively.  Static spherically
symmetric equations are set up, and the three parameters are
interrelated and chosen so that effectively the system reduces to a
one parameter only, $q$. Specific black hole and other type of
one-parameter solutions are studied. First, as a preparation, the
Reissner-Nordstr\"om solution, with $q_1=q_2=q_3=0$, is displayed.
Then, we seek for solutions in which the electric field is regular
everywhere as well as asymptotically Coulombian, and the metric
potentials are regular at the center as well as asymptotically
flat. In this context, the one-parameter model with $q_1\equiv -q$,
$q_2=2q$, $q_3=-q$, called the Gauss-Bonnet model, is analyzed in
detail.  The study is done through the solution of the Abel equation
(the key equation), and the dynamical system associated with the
model. There is extra focus on an exact solution of the model and its
critical properties. Finally, an exactly integrable one-parameter
model, with $q_1\equiv -q$, $q_2=q$, $q_3=0$, is
considered also in detail. A special sub-model, in which the
Fibonacci number appears naturally,
of this one-parameter model is shown, and the corresponding
exact solution is presented.  Interestingly enough, it is a soliton of
the theory, the Fibonacci soliton, without horizons and with a mild
conical singularity at the center.
\end{abstract}
\pacs{04.40.Nr,04.50.Kd,04.70.Bw}
\maketitle

\newpage

\section{Introduction}

A natural and very general extension of the Einstein-Maxwell
Lagrangian yielding a general system of equations for a non-minimal
coupling between the gravitational and electromagnetic fields, with
non-linear terms, was set up and studied in \cite{bl05}. Within this
general theory, a special theory, worth of discussion, arises when one
restricts the general Lagrangian to a Lagrangian that is
Einstein-Hilbert in the gravity term, quadratic in the Maxwell tensor,
and the couplings between the electromagnetism and the metric are
linear in the curvature terms. The motivations for setting up such a
theory are phenomenological, see, e.g., \cite{FaraR,hehl} for reviews
and references.This theory has three coupling constants $q_1$, $q_2$
and $q_3$, which characterize the cross-terms in the Lagrangian
between the Maxwell field $F_{ij}$ and terms linear in the Ricci
scalar $R$, Ricci tensor $R_{ik}$, and Riemann tensor $R_{ikmn}$,
respectively. The coupling constants $q_1$, $q_2$ and $q_3$ have units
of area, and are a priori free parameters, which can acquire specific
values in certain effective field theories.  More specifically, the
action functional of the non-minimal theory linear in the curvature is
\begin{equation}
S=\int\,d^4x\,\sqrt{-g}\,\cal{L}, \label{action1}
\end{equation}
where $g$ is the determinant of the spacetime metric $g_{ik}$, and
the Lagrangian of the theory is
\begin{equation}
{\cal L} = \frac{1}{4\pi}\left(\frac{R}{\kappa} +
\frac{1}{2}\, F_{mn}F^{mn} +
\frac{1}{2} \chi^{ikmn} F_{ik}F_{mn}\right) \,. \label{lagrange1}
\end{equation}
Here $\kappa = 2\,G$, $G$ being the gravitational constant and we are
putting the velocity of light $c$ equal to one, $F_{ik}=\partial_i
A_k-\partial_kA_i$ is the Maxwell tensor, with $A_k$ being the
electromagnetic vector potential, and latin indexes are spacetime
indexes, running from 0 to 3.  The tensor $\chi^{ikmn}$ is the
non-minimal susceptibility tensor given by \begin{equation}
\chi^{ikmn} \equiv \frac{q_1 R}{2}(g^{im}g^{kn} {-} g^{in}g^{km}) {+}
\frac{q_2}{2} (R^{im}g^{kn} {-} R^{in}g^{km} { +} R^{kn}g^{im} {-}
R^{km}g^{in}) {+} q_3 R^{ikmn} \,, \label{susceptibility2}
\end{equation} where $q_1$, $q_2$, and $q_3$ are the mentioned
phenomenological parameters. The action and Lagrangian
(\ref{action1})-(\ref{susceptibility2}) describe thus a
three-parameter class of models, non-minimally coupled, and linear in
the curvature \cite{bl05,FaraR,hehl}.  Lagrangians of this type have
been used and studied by several authors.

The first and important example of a calculation of the three
couplings was based on one-loop corrections to quantum
electrodynamics in curved spacetime, a direct and non-phenomenological
approach considered by Drummond and Hathrell \cite{drummond}. This
model is effectively one-parameter since the coupling constants are
connected by the relations $q_1\equiv-5q$, $q_2=13q$, $q_3=-2q$. The
positive parameter $q$ appears naturally in the theory, and is
constructed by using the fine structure constant $\alpha$, and the
Compton wavelength of the electron $\lambda_{\rm e}$, $q \equiv
\frac{\alpha\lambda^2_{\rm e}}{180\pi}$.  In these models it is useful
to define a radius $r_q$, an effective radius related to the
non-minimal interaction, through $r_q=\sqrt{2|q|}$. Thus, the
corresponding effective radius for the non-minimal interaction in this
case, is the Drummond-Hathrell radius ${r_q}_{\rm DH}$, given by
${r_q}_{\rm DH} \equiv \lambda_{\rm e} \sqrt{\frac{\alpha}{90\pi}}$.
In \cite{Kost1} one also finds a quantum electrodynamics motivation
for the use of generalized Einstein-Maxwell equations.

Phenomenological models, i.e., models based on external considerations
to obtain the couplings, or parameters, $q_1$, $q_2$, and $q_3$, have
also been considered.  Prasanna \cite{Prasanna1,prasannamohanty}
wanting to understand how the strong equivalence principle can be
weakly violated in the context of a non-minimal modification of
Maxwell electrodynamics, has shown that $q_1=q_2=0, q_3\equiv -q$, $q$
a free parameter, is a good phenomenologically model.  Another type of
requirement, one with mathematical and physical motivations, is to
impose that the differential equations forming the non-minimal
Einstein-Maxwell system are of second order (see, e.g.,
\cite{Horn,MHS}).  For instance, in \cite{MHS}, by imposing a
Kaluza-Klein reduction to four dimensions from a Gauss-Bonnet model in
five dimensions, thus guaranteeing second order equations for the
electric field potential $A_i$, and metric $g_{ik}$, it was discussed
a model in which $q_1+q_2+q_3=0$ and $2q_1+q_2=0$, i.e., with
$q_1\equiv -q$, $q_2=2q$ and $q_3=-q$. So the extra non-minimal term
is a kind of Gauss-Bonnet term, and the model is called the
Gauss-Bonnet model.  Yet another type of requirement, this time purely
mathematical, was suggested in \cite{bl05}. The idea is connected with
the symmetries of the non-minimal susceptibility tensor $\chi^{ikmn}$
(see Eq.  (\ref{susceptibility2})).  For instance, one can recover the
relations $q_1+q_2+q_3=0$ and $2q_1+q_2=0$, used in \cite{MHS}, by the
ansatz that the non-minimal susceptibility tensor $\chi_{ikmn}$ is
proportional to the double dual Riemann tensor $^{*}R^{*}_{ikmn}$,
i.e., $\chi_{ikmn} = \gamma\; ^{*}R^{*}_{ikmn}$, for some $\gamma$
(see, \cite{bl05} for details and motivations). Analogously, one can
use the Weyl tensor ${\cal C}_{ikmn}$ in the relation $\chi_{ikmn} =
\omega\,{\cal C}_{ikmn}$, for some $\omega$, or the difference
$R_{ikmn} - {\cal C}_{ikmn}$ instead of $^{*}R^{*}_{ikmn}$, to
introduce some new linear relations between $q_1,q_2,q_3$, namely
$3q_1+q_2=0$ and $q_2+q_3=0$.  Yet another type of requirement is to
choose the parameters so that one obtains exact solutions. As we will
see this will lead to a model with $q_1+q_2+q_3=0$ and $q_3=0$, i.e.,
$q_1\equiv -q$, $q_2=q$, $q_3=0$. Since this model is integrable we
call it the integrable model.  A subcase of this has additional
interest and is called the Fibonacci soliton.

Up to now we have a theory defined through Eqs.
(\ref{action1})-(\ref{susceptibility2}), with each chosen set of
values for the parameters $q_1$, $q_2$, and $q_3$, giving a model. We
have seen that the reduction from three-parameter models to
one-parameter models, specified by the one parameter $q$ and the
relations between $q_1$, $q_2$ and $q_3$, happens in several
instances, either through direct calculation, as in
\cite{drummond,Kost1}, or through phenomenological and other
considerations, as in \cite{Prasanna1}-\cite{MHS} or \cite{bl05} and
here. This certainly simplifies the analysis, and we will consider
this one-parameter type of models, in which $q_1$, $q_2$, and $q_3$,
have a specified relation to the parameter $q$.  For all these models,
one can pick an effective radius $r_q\equiv\sqrt{2|q|}$, as in the
Drummond-Hathrell case, which gives the range of the non-minimal
interaction between the gravitational and electric fields.  Of course,
$r_q$ can be set to zero, in the case the world is pure
Einstein-Maxwell, or otherwise can have a given specified value. The
radius ${r_q}_{\rm DH}$ defined above is a candidate but in principle
not the unique choice. Thus, possible estimations of the parameter
$q$, and so of $r_q$, from, for instance, astrophysical observations,
are undoubtedly of interest (see, e.g., \cite{prasannamohanty}).

Now, after choosing a model, specified by $q$ and by the relations
between $q_1$, $q_2$, and $q_3$, it is important to study exact
solutions.  Exact solutions of the equations of non-minimal
electrodynamics in non-linear gravitational wave backgrounds were
obtained in \cite{B1}--\cite{BL1}, and a non-minimal Bianchi-I
cosmological solution was discussed in \cite{BZ05} in this context.
Here we want to study charged black hole and other charged solutions
of non-minimal models.  The Reissner-Nordstr\"om solution is a
standard solution in pure Einstein-Maxwell theory, with two horizons,
an event and a Cauchy horizon, and a timelike singularity at the
center (see, e.g., \cite{MTW}). Paradigmatic charged black hole
solutions also appear in the framework of Einstein theory minimally
coupled to non-linear electromagnetic fields, as well as other matter
fields. Such solutions were found by Bardeen and others
\cite{Bardeen}-\cite{R9} and the main feature is that they are
regular, without singularities inside the horizon. Within quartic
gravity non-singular charged black hole solutions have also been found
\cite{berej}.  Since the non-minimal theory we are considering
possesses new degrees of freedom, namely, the phenomenological
parameter $q$ and its relations to $q_1$, $q_2$, and $q_3$, we believe
that these allow to introduce new aspects to the problem of finding
black hole and other solutions of each chosen model.  Three aspects
can be mentioned.  First, one wants to have a gauge in order to
compare the new solutions.  Thus, we study the Reissner-Nordstr\"om
solution, the trivial solution in this context, where
$q_1=q_2=q_3\equiv q$ with $q=0$, in order to understand the novel
features, such as causal and singularity structure, of the new
solutions. Second, one should try to search for models exactly or
quasi-exactly soluble. This requirement will take us, among the models
cited above and the many other possible models, to two interesting
models.  They are, the Gauss-Bonnet type model where $q_1\equiv -q$,
$q_2=2q$ and $q_3=-q$, and the integrable model where $q_1\equiv -q$,
$q_2=q$, $q_3=0$.  In both models we perform a detailed
analysis. Third, we vary $q$ within each of the two nontrivial models
and in trying to consider non-minimal extensions of the
Reissner-Nordstr{\"o}m solution, we search for features that are
similar or distinct from the two paradigmatic solutions, the
Reissner-Nordstr{\"o}m solution itself and the Bardeen solutions.  In
our search for charged black hole solutions in non-minimal models we
work in Schwarzschild coordinates and impose certain requirements.
The first requirement is connected with the electric field $E(r)$. We
demand it is a regular function on the interval $0\leq r<\infty$, the
origin being also regular (i.e., the value $E(0)$ is finite), and for
large values of $r$ the electric field is Coulombian, $E(r) \to
\frac{Q}{r^2}$, where $Q$ is the electric charge. The second
requirement is concerned with the metric functions $g_{ik}$. These
should take finite values at the center ($g_{ik}(0) \neq \infty$).
Horizons at $r \neq 0$ are not excluded, and far away the solutions
should be asymptotically flat.  Upon these conditions and solving the
non-minimal equations in certain cases we will find electric charged
solutions with one horizon only, thus causally distinct from the
Reissner-Nordstr\"om.  However, like Reissner-Nordstr\"om, the
solutions have a singularity at the center, although here the
singularity is spacelike instead, as the Schwarzschild case, but
conical, thus much milder.  We have also found in one model a
gravitational charged soliton, without horizons, where the fields are
well behaved, apart from a mild conical singularity at the center.
Although this and other solutions with horizons are almost regular at
the center, we have not obtained strictly non-singular black hole of
the type found by Bardeen and others \cite{Bardeen}-\cite{berej}.  The
difference is based on two aspects. Firstly, we consider spherically
symmetric static solutions with $g_{00}(r) g_{rr}(r) \neq 1$, in
contrast to the Schwarzschild, Reissner-Nordstr{\"o}m, and minimal
regular Bardeen solutions. Secondly, we assume that the values
$g_{00}(0)$ and $g_{rr}(0)$ are finite, but can differ from one.
Moreover, we admit, that $g_{00}(0)$ can, in principle, be equal to
zero. This means that the scalar curvature invariants for such a
metric can take infinite values at the center, and the solution of the
non-minimal Einstein equations is not regular at the center in this
general sense.  However, in many cases the singularity is a conical
one, so much milder than the nasty ones of Schwarzschild and
Reissner-Nordstr\"om.  In summary, we find charged black hole
solutions different in horizon structure from the
Reissner-Nordstr{\"o}m but similar to Bardeen black holes, and
although singular, in certain cases can be considered quasi-regular
conical singularities, in-between Schwarzschild and
Reissner-Nordstr\"om types of singularities and the no singularities
of Bardeen.

This paper is organized as follows. In Section \ref{genericalities},
in particular in subsection \ref{genericalities2}, using the
Lagrangian formalism of the Introduction, we establish a
three-parameter non-minimal Einstein-Maxwell model. In subection
\ref{static}, we set up static equations for studying black holes and
reduce the three-parameter model to a one-parameter model.  In Section
\ref{solutions} we study specific spherically symmetric one-parameter
solutions. In subsection \ref{basics} we define the basic quantitities
and basic variables.  In subsection \ref{RNmodel} we display the
Reissner-Nordstr\"om solution as a preparation, where
$q_1=q_2=q_3\equiv q$ with $q=0$. In subsection \ref{GBmodel} we
analyze in detail a one-parameter model, the Gauss-Bonnet model, with
$q_1\equiv -q$, $q_2=2q$, $q_3=-q$ (i.e., $q_1+q_2+q_3=0$ and
$2q_1+q_2=0$), using the known solution of the Abel equation, the key
equation of the model, and the dynamical system associated with this
model.  We display the charged black hole solutions, and we focus on a
specific exact solution and its critical properties. In subsection
\ref{nonamemodel} we consider in detail an exactly integrable
one-parameter model, the integrable model, with $q_1\equiv -q$,
$q_2=q$, $q_3=0$ (i.e., $q_1+q_2+q_3=0$ and $q_3=0$). We examine a
special sub-model, the Fibonacci soliton, of this one-parameter model
and present the corresponding exact solution.  In subsection
\ref{table} we present, by means of a table, the summary of the
results of the models studied.  In Section \ref{conc} we conclude.

\section{Non-minimal coupling, linear in the curvature, between
gravity and electromagnetism: Equations and reduction from three
parameters to one parameter for static spherically symmetric systems}
\label{genericalities}

\subsection{The three-parameter model:
general equations}
\label{genericalities2}

The variation of the Lagrangian
(\ref{action1})-(\ref{susceptibility2}) with respect to the metric
yields (see, \cite{bl05} for details),
\begin{equation}
R_{ik} - \frac{1}{2} R \ g_{ik} = \kappa \left[ T^{(0)}_{ik} + q_1
T^{(1)}_{ik} + q_2 T^{(2)}_{ik} + q_3 T^{(3)}_{ik} \right] \,.
\label{standardform}
\end{equation}
The energy-momentum tensor of the pure electromagnetic field,
$T^{(0)}_{ik}$, is
\begin{equation}
T^{(0)}_{ik} = \frac{1}{4} g_{ik} F_{mn}F^{mn} - F_{im} F_{k}^{\
m}\,. \label{T0}
\end{equation}
The definitions for the other three parts of the stress-energy
tensor, $T^{(1)}_{ik}$, $T^{(2)}_{ik}$ and $T^{(3)}_{ik}$, are
\begin{equation}
T^{(1)}_{ik} = R \ T^{(0)}_{ik} - \frac{1}{2} R_{ik} F_{mn}F^{mn}
- \frac{1}{2} g_{ik} \nabla^l \nabla_l (F_{mn}F^{mn}) +
\frac{1}{2} \nabla_{i} \nabla_{k} (F_{mn}F^{mn})  \,,
\label{part1}
\end{equation}
\begin{eqnarray}
T^{(2)}_{ik} &=& - \frac{1}{2}g_{ik}\left[ \nabla_{m}
\nabla_{l}(F^{mn}F^{l}_{\ n} ) - R_{lm}F^{mn}F^{l}_{\
n}\right] - F^{ln}(R_{il}F_{kn} + R_{kl}F_{in}) -
\nonumber\\
&& {-} R^{mn} F_{im} F_{kn} {-} \frac{1}{2} \nabla^l \nabla_l
(F_{in}F_{k}^{\ n}) {+} \frac{1}{2}\nabla_l \left[
\nabla_i(F_{kn}F^{ln}) {+} \nabla_k(F_{in}F^{ln}) \right] \,,
\label{part2}
\end{eqnarray}
\begin{equation}
T^{(3)}_{ik} = \frac{1}{4}g_{ik} R^{mnls}F_{mn}F_{ls} {-}
\frac{3}{4}F^{ls}(F_{i}^{\ n}R_{knls}+F_{k}^{\ n}R_{inls})
{-} \frac{1}{2}\nabla_{m} \nabla_{n}(F_{i}^{\ n}F_{k}^{\
m} {+} F_{k}^{\ n}F_{i}^{\ m})\,. \label{part3}
\end{equation}
In addition, the non-minimal electrodynamics associated with the
Lagrangian (\ref{action1})-(\ref{susceptibility2}) obeys the equation
\begin{equation}
\nabla_k \left[ F^{ik} + \chi^{ikmn} F_{mn} \right] = 0 \,, \quad
\nabla_k F^{*ik} = 0 \,, \label{m1}
\end{equation}
where $F_{mn}$ is the Maxwell tensor and $F^{*}_{kl}$ is dual to
it. Consider now these master equations in the case of a static
spherically symmetric spacetime.

\subsection{Static spherically symmetric non-minimally coupled
fi\-elds. Reduced three-parameter system of equations: one-pa\-rameter
models}
\label{static}

\subsubsection{Preliminaries}

Using Schwarzschild coordinates, the line element for a static
spherically symmetric system can be put in the form
\begin{equation}
ds^2 = B(r) \ c^2 dt^2 - A(r) \ dr^2 - r^2\,(d\theta^2 + \
\sin^2\theta \ d\varphi^2) \,,
\label{metric1}
\end{equation}
where $B$ and $A$ are metric potentials that depend on the radial
coordinate $r$ only. This form of the line element is useful as when
$B$ and $1/A$ are simultaneously zero it signals the presence of an
event horizon.  Assume also that the electromagnetic field inherits the
static and spherical symmetries. Then the potential four-vector of
the electric field $A_i$ has the form
\begin{equation}
A_i = A_0(r) \delta^0_i\,.
\label{potentialA}
\end{equation}
From (\ref{potentialA}) the Maxwell tensor is equal to $F_{ik}=
A^{\prime}_0(r) (\delta^{r}_{i}\delta^{0}_{k} -
\delta^{0}_{i}\delta^{r}_{k})$, where a prime denotes the derivative
with respect to $r$.  To characterize the electric field, it is useful
to introduce a new scalar quantity $E(r)$ as $E^2(r) \equiv -
\frac{1}{2} F_{mn}F^{mn}$. Then the electric field squared is
$E^2(r)=\frac{1}{AB}(A^{\prime}_0)^2\,$ from which one obtains in turn
$F_{r0}= - (AB)^{\frac{1}{2}} E(r)$.  Since the expressions $1/A$ and
$\sqrt{AB}$ enter frequently in the master equations, it is sometimes
convenient to use the functions $N(r)$ and $\sigma(r)$ defined as
$N(r) \equiv
\frac{1}{A(r)}$ and $\sigma(r)\equiv\sqrt{A(r)B(r)}$. In summary, the
functions
\begin{equation}
E^2(r) \equiv \frac{1}{\sigma^2}(A^{\prime}_0)^2 \,,
\label{EsigmaN}
\end{equation}
and
\begin{equation}
N(r) \equiv
\frac{1}{A(r)} \,, \quad \sigma(r) \equiv \sqrt{A(r)B(r)} \,,
\label{sigmaN}
\end{equation}
are alternatives to the functions $A_0(r)$, $A(r)$, and $B(r)$.

\subsubsection{Key equation for the Maxwell field and its solution}

The Maxwell equations (\ref{m1}) with (\ref{potentialA})
give only one non-trivial
equation, namely,
\begin{equation}
\left[r^2 E(r) \left(1 + 2 \chi^{0r}_{\ \ 0r}(r) \right)
\right]^{\prime} = 0 \,, \label{Eequ}
\end{equation}
which can be integrated immediately to give
$$
E(r) \left\{ r^2 \left[ 1 + (q_1+q_2+q_3)\left( N^{\prime \prime}
+ 3N^{\prime} \ \frac{\sigma^{\prime}}{\sigma} + 2N
\frac{\sigma^{\prime \prime}}{\sigma} \right) \right] \right.
$$
\begin{equation}
\left. + 2r (2q_1+q_2) \left( N^{\prime} + N
\frac{\sigma^{\prime}}{\sigma}\right) + 2q_1 (N -1) \right\}= Q
\,, \label{1e}
\end{equation}
where $Q$ is a constant, to be associated
with the central electrical charge of the solution.
This equation gives the electric field of
a central charge, corrected by the radial component of the
dielectric permeability tensor, $1 + 2 \chi^{0r}_{ \ \ 0r}(r)$. This
component
describes the vacuum screening effect on the central charge, due
to the interaction of the vacuum with curvature, analogously to
the screening of a charge by a non-homogeneous medium in a
spherical cavity. Supposing the spacetime to be asymptotically
flat, i.e., $R^i_{ \ klm}(\infty) = 0$, one can see that
(\ref{1e}) yields asymptotically the Coulomb law $E \to Q/r^2$,
and the constant $Q$ indeed coincides with the total
electric charge of
the object.

\subsubsection{Key equations for the gravitational field}

The equations for the gravitational field (\ref{standardform})
with (\ref{T0})-(\ref{part3}) and the metric potentials (\ref{metric1})
redefined as in
(\ref{sigmaN}) can be rewritten as a pair of equations for
$N(r)$ and $\sigma(r)$, respectively,
\begin{eqnarray}
\frac{[r(1-N)]^{\prime}}{\kappa r^2} = -  \left( E^2\right)^{\prime
\prime} N (q_1+q_2+q_3)\nonumber\\
+ \left( E^2\right)^{\prime} \left[-
\frac{1}{2}(q_1+q_2+q_3)\left( N^{\prime} + \frac{8N}{r}\right)+
\frac{N}{r}(2q_1+q_2) \right]\nonumber\\
+ E^2 \left[ \frac{1}{2} + (q_1+q_2+q_3)\left( N^{\prime \prime} +
3 N^{\prime} \ \frac{\sigma^{\prime}}{\sigma} + 2N
\frac{\sigma^{\prime \prime}}{\sigma} - \frac{N^{\prime}}{r} - 2
\frac{N}{r^2} \right) \right.\nonumber\\
\left. + (2q_1+q_2)\left( 2\frac{N^{\prime}}{r} + 2\frac{N}{r}
\frac{\sigma^{\prime}}{\sigma} + \frac{N}{r^2} \right) + q_1
\frac{(N-1)}{r^2} \right]\,, \label{2e}
\end{eqnarray}
\begin{eqnarray}
\frac{2\sigma^{\prime}}{\kappa r \sigma} = -  \left(
E^2\right)^{\prime \prime} (q_1+q_2+q_3) +\left(
E^2\right)^{\prime} \left[(q_1+q_2+q_3) \left(
\frac{\sigma^{\prime}}{\sigma} - \frac{4}{r}\right) + (2q_1+q_2)
\frac{2}{r}  \right]\nonumber\\
+ E^2 \left[(q_1+q_2+q_3) \frac{2\sigma^{\prime}}{ r \sigma} -
\frac{2q_3}{r^2} \right] \,.\label{3e}
\end{eqnarray}
The first equation can be reduced to an equation for $E(r)$ and
$N(r)$, by extracting the term $\frac{\sigma^{\prime}}{\sigma}$ from
the second one.  The second equation contains the unknown functions
$E(r)$ and $\sigma(r)$ only. Thus, Eqs. (\ref{1e})-(\ref{3e}) form the
key system of equations for the non-minimal Einstein-Maxwell model of
a static spherically symmetric object. It is a system of three
ordinary differential equations of second order for the three unknown
functions, $E(r)$, $N(r)$ and $\sigma(r)$. The form of equations is
not canonical. In principle, the electric field $E(r)$ can be
extracted explicitly from (\ref{1e}) as a function of $N^{\prime
\prime}$, $N^{\prime}$, $N$, $\frac{\sigma^{\prime \prime}}{\sigma}$,
$\frac{\sigma^{\prime}}{\sigma}$ and $r$. Inserting such $E(r)$ into
the Eqs. (\ref{2e})-(\ref{3e}), we obtain equations for $N(r)$
and $\sigma(r)$ of fourth order in their derivatives.

\subsubsection{General features and notes}

Below we focus on models admitting solutions to Eqs.
(\ref{1e})-(\ref{3e}), such that they can be represented by a series
expansion regular at $r=0$, i.e.,
\begin{equation} {\cal A}(r \to 0) =
{\cal A}(0) + {\cal A}^{\prime}(0) \ r + \frac{1}{2}{\cal
A}^{\prime\prime}(0) \ r^2 + ... \, \label{regular}
\end{equation}
where ${\cal A}(r)$ symbolizes generically the functions $E(r)$,
$N(r)$, and $\sigma(r)$.  Our purpose is to find solutions satisfying
three conditions: First, the electric field $E(r)$ should be a
continuous function regular at $r=0$ ($E(0) \neq \infty$) and also
should be of Coulombian form at $r \to \infty$. Second, the metric
functions $N(r)$ and $\sigma^2(r) N(r)$ should be regular at
$r=0$. Third, in terms of the functions $A(r)$ and $B(r)$ the asymptotic
flatness requires that $B^{\prime \prime}(\infty) = A^{\prime
\prime}(\infty) = B^{\prime}(\infty) A^{\prime}(\infty)=0$, and
$A(\infty)={\rm const}$, $B(\infty) = {\rm const}$.  So, essentially,
one can put, $\sigma(\infty)=1$ and $N(\infty)=1$.

Note that the regularity of the functions $E(r)$, $N(r)$, and
$\sigma(r)$ at $r=0$ does not guarantee that the solution of the
Einstein-Maxwell model is characterized by regular curvature
invariants. For instance, when $N(0)$ is finite but $N(0) \neq 1$, the
model displays a conical singularity and the scalar invariants of the
curvature tend to infinity as $r \to 0$. In considering solutions such
that the fields are finite we try to be as close as possible to
Bardeen's idea of having black hole solutions without singularities,
by finding a regular $E(r)$ and putting the metric coefficients in the
form $N(r) = 1 -2Mr^2 (r^2+r^2_0)^{-\frac{3}{2}}$ and $\sigma(r)=1$,
for some $r_0$ \cite{Bardeen}.  As we will see it will turn out that
this is not achieved, since although the electric and metric
potentials are regular, the black hole solutions found here are
singular at the center, where the curvature invariants blow up.
Notwithstanding, these solutions are very interesting.  Using
the ansatz (\ref{regular}) and the Eqs. (\ref{1e})-(\ref{3e}) one
can couple the values $E(0)$, $N(0)$, $\sigma(0)$, and $q_1$, $q_2$,
$q_3$. The relations are different for the cases $\sigma(0) = 0$
and $\sigma(0) \neq 0$, which we now analyze.

\vspace{2mm}
\noindent {\it (i) $\sigma(0) = 0$:}
\noindent When $\sigma(0) = 0$, but $\sigma^{\prime}(0) \neq 0$,
one obtains from the system (\ref{1e})-(\ref{3e}) that
$\left(\frac{\sigma^{\prime}}{\sigma}\right)(r
\to 0) \to \frac{1}{r}$, and the decomposition (\ref{regular}) is
valid at $r \to 0$, when the following conditions are satisfied,
\begin{eqnarray}
E(0) \frac{3q_1 q_2 + q_2^2 + 2q_1q_3}{3q_1+q_2-q_3} =Q\,, \quad
      E^2(0) (q_1 + q_2)-1=0 \,,
\nonumber\\
N(0) \left[2(3q_1 + q_2 - q_3)\right]= 2q_1 +
q_2\,, \label{regularity2}
\end{eqnarray}
for generic $q_1$, $q_2$, and $q_3$. There are two specific cases.
When $q_1+q_2+q_3=0$, but both $2q_1+q_2 \neq 0$ and $q_1+q_2 \neq 0$,
then
$Q=\frac{1}{2} E(0) (q_2-q_1)$ and $N(0)=\frac{1}{4}$. When
$q_1+q_2+q_3=0$ and $2q_1+q_2 = 0$, simultaneously, then $\kappa q_1
E^2(0)=-1$, providing $q_1$ is negative, and $N(0)$ is fixed by the
relation $N(0)= 1 + \frac{Q}{2q_1 E(0)} \neq 1$.  As in the
case $\sigma(0) \neq 0$, see below, here
the Ricci scalar $R(r)$ takes an infinite value at $r=0$.

\vspace{2mm}
\noindent {\it (ii)  $\sigma(0) \neq 0$:}
\noindent When  all three functions, $E(r)$, $N(r)$, and $\sigma(r)$
are regular at $r=0$, and $\sigma(r)$, appearing in the denominator
of Eqs. (\ref{1e})-(\ref{3e}), does not vanish at $r=0$,
one obtains from the system (\ref{1e})-(\ref{3e})
the following set of equations
\begin{eqnarray}
E(0) \ 2 q_1 [N(0) - 1] = Q \,, \quad E^2(0) \ 2 q_3 = 0 \,,
\nonumber\\
N(0) \left[ 1 + \kappa E^2(0) (q_1 - q_2 - 2 q_3) \right] = 1 +
q_1 \kappa E^2(0) \,, \label{regularity1}
\end{eqnarray}
for generic $q_1$, $q_2$, and $q_3$.
Since the charge of the object, $Q$, is considered to be
non-vanishing, one obtains immediately from the first equation
of the set that $E(0) \neq 0$
and $N(0) \neq 1$. Thus, we infer,
\begin{equation}
q_1 = \frac{Q}{2E(0) [N(0) - 1]} \,, \quad q_2 =  \frac{2 [N(0)-1]
+ \kappa E(0) Q}{2 N(0) \kappa E^2(0)} \,, \quad q_3=0\,. \label{r2}
\end{equation}
The relations (\ref{regularity1}) give that the curvature invariants
are infinite in the center $r=0$. For instance, when $N(r)$ and
$\sigma(r)$ are regular in the center and $\sigma(0)\neq 0$, then the
Ricci scalar
\begin{equation}
R(r) = 2 N \frac{\sigma^{\prime \prime}}{\sigma} + N^{\prime \prime}
+ 3N^{\prime} \frac{\sigma^{\prime}}{\sigma} +
\frac{4}{r} \left(N^{\prime} + N \frac{\sigma^{\prime}}{\sigma} \right)
+ \frac{2}{r^2} (N-1)
\label{Ricci}
\end{equation}
tends to infinity at $r \to 0$, since $N(0) \neq 1$, as well as generally
$N^{\prime}(0) \neq - N(0) \frac{\sigma^{\prime}}{\sigma}(0)$.

\subsubsection{The order of differential equations and the
choice of the parameters: one-parameter
models}

Now we want analyze the simplest cases of the system
of Eqs. (\ref{1e})-(\ref{3e}). One sees
that there is an immediate simplification when
$q_1+q_2+q_3=0$, since second order derivatives and products
of first order derivatives disappear from the equations.
In such a case the system (\ref{1e})-(\ref{3e}) reduces to
\begin{equation}
E(r) \left\{ r^2 + 2r (2q_1+q_2) \left( N^{\prime} + N
\frac{\sigma^{\prime}}{\sigma}\right) + 2q_1 (N -1) \right\}= Q
\,, \label{11e}
\end{equation}
\begin{equation}
\frac{r \sigma^{\prime}}{\kappa \sigma} = r (2q_1+q_2)
\left(E^2\right)^{\prime} - q_3 E^2  \,,
\label{12e}
\end{equation}
\begin{eqnarray}
\frac{[r(1-N)]^{\prime}}{\kappa} = r \left( E^2\right)^{\prime}
(2q_1+q_2) N +\nonumber\\
+ E^2 \left[ \frac{r^2}{2} - q_1 + 2 r (2q_1+q_2)\left(
N^{\prime} + N \frac{\sigma^{\prime}}{\sigma} \right) + N (3q_1
+q_2) \right]\,. \label{13e}
\end{eqnarray}
Moreover, the three-dimensional matrix, composed of the
coefficients in front of the first derivatives $(E^2)^{\prime}$,
$N^{\prime}$ and $\sigma^{\prime}$, has rank two. This means
that for $q_1+q_2+q_3=0$ the system (\ref{11e})-(\ref{13e}) can be
reduced to one algebraic equation and two differential equations
of the first order. The corresponding algebraic equation is
\begin{equation}
\kappa (2q_1+q_2)(r^2-2q_1) E^3  - 2 \kappa (2q_1+q_2)Q E^2 +
[r^2 + 2q_3(N-1)] E - Q = 0 \,,
\label{cubic}
\end{equation}
which in turn links two functions, $E(r)$ and $N(r)$.
Now from equation (\ref{cubic}), one can consider three
subcases, that emerge from the case $q_1+q_2+q_3=0$.
First we consider briefly the trivial case in this context,
$
q_1=q_2=q_3\equiv q, \; q=0,
$
i.e., the Reissner-Nordstr{\"o}m limit, see subsection \ref{RNmodel}.
Then we consider two interesting non-trivial
cases: first, $q_3 \neq 0$, second, $q_3 = 0$,
When $q_3 \neq 0$ it is easy to
express $N(r)$ in terms of $E(r)$. In the subsection \ref{GBmodel}
we consider
a specific model in this class, characterized by the supplementary
condition
$
q_1\equiv -q, \,q_2=2q, \,q_3=-q\;\;
({\rm i.e.}, \,q_1+q_2+q_3=0,\, 2q_1+q_2=0)\,.
$
This is the Gauss-Bonnet type model, which has
been considered as an important model in \cite{Horn,MHS},
and for which we present an extended analysis with
relevant new details.
For the second case
$
q_1\equiv -q, \,q_2=q, \,q_3=0\;\;
({\rm i.e.}, \,q_1+q_2+q_3=0,\,q_3=0)\,,
$
$E(r)$ decouples from $N(r)$ and we deal with a cubic equation for
the determination of the electric field. We will consider such a
model, the integrable model, in subsection \ref{nonamemodel}.

\section{Solutions of the reduced three-parameter model: solutions of
one-parameter models}
\label{solutions}

\subsection{Basic quantities and variables}
\label{basics}
One should first define
three quantities, $r_M$, $r_Q$, $E_Q$, as follows
\begin{equation}
r_M\equiv 2G\,M\,,\quad
r_Q \equiv \sqrt{G} |Q| \,, \quad
E_Q \equiv \frac{Q}{r^2_Q} \,.\quad
\label{rq}
\end{equation}
Now, the models we are going to discuss here are essentially
one parametric, with
$q_1$, $q_2$, and $q_3$ being a multiple of some parameter $q$.
It is useful to introduce first a quantity $r_q$, given through
\begin{equation}
r_q = \sqrt{2|q|}\, ,\; {\rm and} \quad 2q = \pm r^2_q \,,
\label{rqpropersaid}
\end{equation}
with $r_q$ being a radius.
From $r_M$, $r_Q$, and $r_q$, one can then construct two independent
dimensionless quantities, namely
\begin{equation}
a\equiv \frac{2q}{r^2_Q}= \pm \frac{r^2_q}{r^2_Q} \,,
\label{aratio}
\end{equation}
and
\begin{equation}
K\equiv\frac{r_M}{r_Q} \,.
\label{K0}
\end{equation}
The $a$ quantity gives the deviation from the standard
Reissner-Nordstr\"om case, and $K$ fixes the ratio between the total
mass to the total charge of the object.

In addition, for what follows below, it is  useful to write the equations
of motion by
defining two dimensionless variables, a normalized radius $x$ and a
normalized electric field $Z(x)$, defined as follows
\begin{equation}
x= \frac{r}{r_Q}\,, \quad Z(x)=\frac{E(r)}{E_Q} \,. \label{dimssRN}
\end{equation}
Also, the function $N(x)$ can be defined in terms of another
useful function $y(x)$, i.e.,
\begin{equation}
N(x)=1-\frac{y(x)}{x} \,.
\label{NRN}
\end{equation}
The physical interpretation of $y(x)$ is
connected with the effective mass of the object, i.e., $M(r)$,
as we will see below.
With these quantities defined we now discuss the
Reissner-Nordstr{\"o}m limit and the two new models.

\subsection{The Reissner-Nordstr{\"o}m limit:
$q_1=q_2=q_3\equiv q$, with $q=0$}
\label{RNmodel}

When the non-minimal parameters $q_1$, $q_2$, and $q_3$ are set to
zero, i.e.,
\begin{equation}
q_1=q_2=q_3\equiv q\,, \quad q=0\,,
\label{qsRN}
\end{equation}
and so
$a=0$ as well, we can integrate immediately Eqs. (\ref{1e})-(\ref{3e}).
In terms of the above functions,
the system of key equations can be rewritten as
\begin{equation}
x^2 Z -1 = 0 \,,
\label{1eqdimRN}
\end{equation}
\begin{equation}
x N^{\prime}(x) + N(x)= 1 -  2 Z + x^2 Z^2 \,,
\label{2eqdimRN}
\end{equation}
\begin{equation}
\frac{\sigma^{\prime}(x)}{\sigma} = 0 \,.
\label{3eqdiRNm}
\end{equation}
The solutions to these equations are
\begin{equation}
Z(x)=\frac{1}{x^2} \,,
\label{electricRN}
\end{equation}
\begin{equation}
N(x)=1-\frac{K-1/x}{x} \,,
\label{NRNnew}
\end{equation}
\begin{equation}
\sigma(x)=1 \,.
\label{sigmaRN}
\end{equation}
The function $y(x)$ in (\ref{NRN}) is here given by
\begin{equation}
y(x)=K-\frac{1}{x} \,,
\label{yRN}
\end{equation}
where $K$ is defined in (\ref{K0}), and
one can see through Eq. (\ref{2eqdimRN}) that $y$ obeys
the equation
\begin{equation}
\frac{dy}{dx}=\frac{1}{x^2} \,.
\label{ydashRN}
\end{equation}
Of course one can transform to the original fields
$E(r)$, $A(r)$, and $B(r)$, giving
\begin{equation}
E(r) = \frac{Q}{r^2} \,,
\label{RN}
\end{equation}
\begin{equation}
\frac{1}{A(r)} =1 - \frac{{2GM(r)}}{r}\,,
\label{ARN}
\end{equation}
\begin{equation}
B(r) = 1 - \frac{{2GM(r)}}{r} \,,
\label{BRN}
\end{equation}
which are the usual Reissner-Nordstr\"om functions.  Note, then, that
the physical interpretation of $y(x)$ given in (\ref{yRN}) is
connected with the effective mass of the object, i.e., $M(r)$, which
in turn is given by the definition
\begin{equation}
N(r)=\frac{1}{A(r)}= 1 - \frac{2 G M(r)}{r} = 1 - \frac{y(x)}{x}\,.
\label{NYXRN}
\end{equation}
Thus, $y(x)$ is related to
the dimensionless effective mass, since
$y(x) = \frac{r_{M}}{r_{Q}} \frac{M(r)}{M}$,
with $M(r)\equiv M-\frac{Q^2}{2r}$,
$M$ being the asymptotic mass of the object.

We study in this context the usual static spacetimes with $a=0$, i.e.,
the Schwarzschild and Reissner-Nordstr\"om spacetimes, which are
special solutions of the full system of equations. These two solutions
are heavily singular, both the metric and the Kretschmann scalar
diverge at $r=0$. So these solutions are outside the spirit of the
solutions we want to find. They do not serve as models.
However, they are of interest to set the
nomenclature, and to have a gauge with which we can compare the
solutions we find in the two models studied below.  Note that for these
solutions $\sigma^2(x)= A(r)\,B(r)=1$. In these cases the problem of
searching for horizons is equivalent to finding their radial position
$r_{\rm h}$ through the solutions of $B(r) = \frac{1}{A(r)} \equiv 1 -
\frac{2GM(r)}{r} = 0$, where $M(r)$ is the effective mass. In terms of
dimensionless mass $y(x)$, see Eq. (\ref{NYXRN}), or Eq. (\ref{yRN}),
this condition can be written as $y(x)=x$. For the Schwarzschild
metric $y(x)=K$, where $K=r_M/r_Q$ is defined in (\ref{K0}). Then, one
obtains only one horizon at $x=x_{\rm h} = K$, which is just the
Schwarzschild radius, $r=r_M$.  For the Reissner-Nordstr\"om metric one
has $y(x)= K - \frac{1}{x}$, and the equation $K - \frac{1}{x} = x$
gives three different cases:
(i) $K>2$ (i.e., $r_M>2r_Q$, or $GM^2>Q^2$ in the standard
notation): there are two solutions $x_{\rm h}= \frac{K}{2} +
\sqrt{\frac{K^2}{4} -1}$, and $x_{\rm h}= \frac{K}{2}
-\sqrt{\frac{K^2}{4} -1}$, corresponding to the outer and inner
horizons, respectively, of a usual Reissner-Nordstr\"om black hole.
(ii) $K=2$ (i.e., $r_M = 2 r_Q$, or $GM^2=Q^2$):
there is one horizon only, given by $x_{\rm h}=1$, corresponding
to an extremal black hole.
(iii) $K<2$ (i.e., $r_M < 2 r_Q$, or $GM^2<Q^2$):
there is no solution to the equation $y(x)=x$.
The object is a naked singularity.

\subsection{The Gauss-Bonnet model:
$q_1\equiv -q$, $q_2=2q$, $q_3=-q$
(i.e., $q_1+q_2+q_3=0$ and $2q_1+q_2=0$)}
\label{GBmodel}

\subsubsection{Preliminaries}

Consider now, in Eqs.
(\ref{action1})-(\ref{susceptibility2}), the following specific
one-parameter model
\begin{equation}
q_1\equiv -q, \,q_2=2q, \,q_3=-q\;\;
({\rm i.e.}, \,q_1+q_2+q_3=0,\, 2q_1+q_2=0)\,,
\label{modeladopted}
\end{equation}
for some parameter $q$. In this model the susceptibility tensor is
proportional to the double-dual Riemann tensor and is divergence-free
\cite{bl05}, i.e., $ \chi_{ikmn} = q \; ^{*}R^{*}_{ikmn}$, and
$\nabla_n \chi^{ikmn} = 0 \,.  \label{susceptibility_Riemann5} $
Moreover, the coupled Einstein and electromagnetic equations are
second order in the derivatives, which is the reason why this model is
called a Gauss-Bonnet model. Gauss-Bonnet gravity in five and higher
dimensions has the property that its equations are of second order,
Lovelock gravity being a generalization of it. Actually, one can show
that the model specified by (\ref{modeladopted}) comes from
Kaluza-Klein reduction to four dimensions from five a dimensional
Gauss-Bonnet theory, i.e., Einstein gravity plus a Gauss-Bonnet term
\cite{MHS}.
Using (\ref{modeladopted}), Eqs. (\ref{11e})-(\ref{13e}) convert,
respectively, into
\begin{equation}
E(r) \left\{ r^2 + 2q (1-N) \right\}= Q
\,, \label{11e1}
\end{equation}
\begin{equation}
r \ \frac{\sigma^{\prime}}{\sigma} = \kappa q E^2  \,,
\label{12e11}
\end{equation}
\begin{equation}
[r(1-N)]^{\prime} = \frac{1}{2}\kappa E^2 \left[ r^2 + 2 q(1 - N)
\right]\,.
\label{13e1}
\end{equation}
After appropriate redefinitions these equations agree with the
ones discussed in \cite{MHS}. The correspondingly modified algebraic
Eq. (\ref{cubic}) coincides with (\ref{11e1}). We now consider
two ways of analyzing this system of equations: first, we use a power
series expansion, and second, we apply the formalism of dynamical
systems.

\subsubsection{Abel equation and its solution}

\noindent {(I) The Abel equation}
\vskip 0.1cm
\noindent
Using from Eq. (\ref{11e1}) that
\begin{equation}
N(r) = 1 + \frac{1}{2q}\left[ r^2 - \frac{Q}{E(r)} \right]\,,
\label{equaNE}
\end{equation}
one obtains the Abel equation (see, e.g., \cite{PolZa})
for $E(r)$ from (\ref{13e1}),
\begin{equation}
r E^{\prime}(r) = E - 3 \frac{r^2}{Q} E^2 - \kappa q E^3 \,.
\label{equaE}
\end{equation}
Similarly, eliminating $E(r)$ from Eq. (\ref{11e1}) one can transform
Eq. (\ref{13e1}) into the
Abel equation for a function $\Theta(r)$,
here defined as
\begin{equation}
\Theta(r)\equiv 1-N(r)\,.
\end{equation}
Thus, Eq. (\ref{13e1}) is given by
\begin{equation}
\left[ r \Theta^{\prime}(r)  + \Theta(r) \right] \left[ r^2 + 2q \Theta(r)
\right]  =
\frac{\kappa Q^2}{2}  \,. \label{equaT}
\end{equation}
Clearly, the values $E(0)$ and $N(0)$ are related through $N(0) = 1 -
\frac{Q}{2q E(0)}$. Searching for solutions $N(r)$ regular at $r=0$,
we have to consider $E(0)$ to be non-vanishing, $E(0) \neq 0$. In such
a case Eq. (\ref{equaE}) yields $\kappa q E^2(0) = 1$. This means that $q$
has to be positive in these solutions.  Thus, although
we analyze all cases, we tend to focus in models with
$a \equiv \frac{2q}{r^2_Q}>0$.  Finally, a possible solution, regular
at $r=0$, should be characterized by $\sigma(0) =0$.  It is convenient
to use the auxiliary quantities $r_q \equiv\sqrt{2|q|}$, $r_{Q}$, and
$a$ used before (see Eqs. (\ref{rqpropersaid})-(\ref{aratio})),
to write
\begin{equation}
E(0) = \frac{1}{\sqrt{\kappa q}}
= \frac{Q}{r_Q r_q} = \frac{E_{Q}}{\sqrt{a}}
\,, \quad
N(0) = 1 - \frac{Q}{2} \sqrt{\frac{\kappa}{q}} = 1- \frac{r_Q}{r_q} = 1-
\frac{1}{\sqrt{a}} \,.
\label{equaEN}
\end{equation}
Using Eqs. (\ref{11e1})-(\ref{13e1}), plus the asymptotic
conditions $\sigma(\infty)=1$, $N(\infty)=1$, as well as the condition
that the electric field is asymptotically Coulombian $E \to Q/r^2$,
one can obtain the following formula for the asymptotic mass $M$,
\begin{equation}
r_{M} = 2GM = \lim_{r\to\infty} \left\{\sigma^2
\left[r(1-N) -\frac{1}{2} r\kappa Q E +
2 N \kappa q E^2 \right] \right\} = \lim_{r\to\infty} r(1-N)  \,.
\label{equaM2}
\end{equation}

\vskip 2.7cm
\noindent {(II) The solution of the Abel equation for small $r$}
\vskip 0.1cm
\noindent
In the vicinity of the point $r=0$ the solutions for $E(r)$, $N(r)$
and $\sigma(r)$ are assumed to have a polynomial form of the type
given in Eq. (\ref{regular}). The decomposition of a regular
solution $E(r)$ with non-vanishing $E(0)$ is
\begin{equation}
E(r) \rightarrow  \frac{E_{Q}}{\sqrt{|a|}}\left[1-\frac{3}{\sqrt{|a|}}
\left(\frac{r}{r_Q}\right)^2 + ... \right] \,,
\label{small1}
\end{equation}
and the corresponding $N(r)$ with finite $N(0)$ is given by
\begin{equation}
N(r)  \rightarrow  1  -  \frac{1}{\sqrt{|a|}} + \frac{1}{4 a^2}
\left(\frac{r}{r_Q}\right)^2 + ... \,.
\label{small2}
\end{equation}
For this solution the effective mass $M(r)$ becomes equal to zero at
$r=0$.
Moreover, $N(0)= 0$, when electric and non-minimal radii coincide,
i.e., $r_Q=r_q$, $a=1$.

\begin{figure}
\centerline{\includegraphics[width=6.76in,height=5.01in]{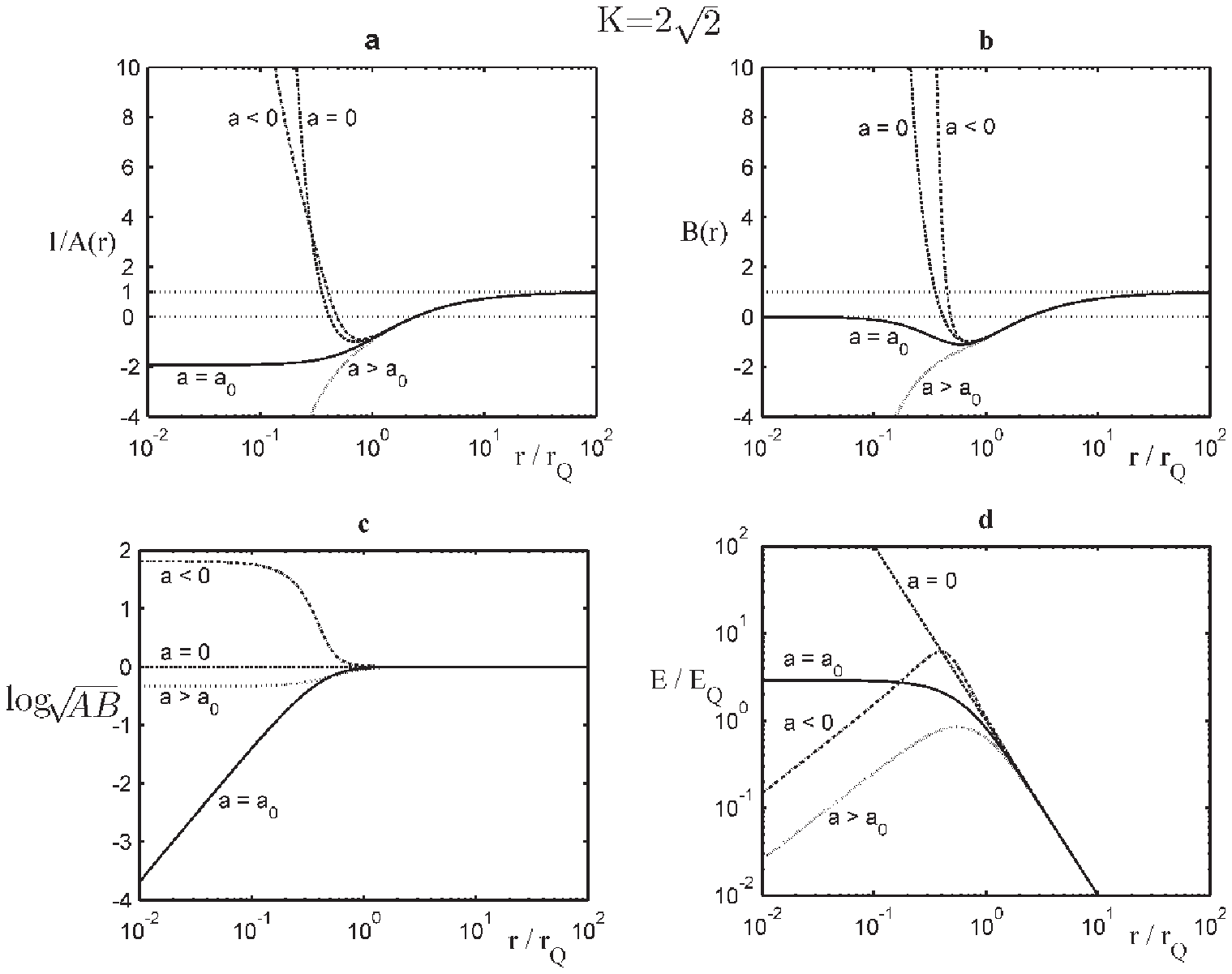}}
\caption {{\small Non-minimal solution of the Gauss-Bonnet
model, with $q_1=-q$, $q_2=2q$
and $q_3=-q$, of gravitational electrically charged objects
characterized by $a=2q/r_Q^2$, for $K=2\sqrt2$ ($K>2$) --}
{\small Plots (a), (b), (c), and (d) depict the metric potentials
$1/A(r)$ and $B(r)$, the function $\log{\sqrt{A(r)B(r)}}$, and the
electric field $E(r)/E_Q$, respectively, as functions of
$x=\frac{r}{r_Q}$, for solutions with different values of the
non-minimal quantity $a$. The Reissner-Nordstr\"om black
hole has $a=0$, the curves of which are clearly shown in the plots. In
this case the curves for $1/A(r)$ and $B(r)$ have two zeros
representing the inner and outer horizons, and for $r\to\infty$ they
go to one, while the curve $E(r)/E_Q$ tends to zero, respectively, and
when $r\to0$ these curves tend to $-\infty$ (the electric field is in
a logarithmic scale).  For $a<0$ and $0<a<a_0$ the black holes behave
quite similarly as the case $a=0$, with two horizons, and the function
$E(r)/E_Q$ tends to finite values as $r\to0$. For $a=a_0$ the curve
for $1/A(r)$ tends to a finite negative value when $r \to 0$, and
takes the value zero only once. On the other hand $B(r)$ has two zeros
one at the same point as $1/A(r)$, the event horizon, and the other at
$r=0$, signaling the presence of a singularity there.  The function
$E(r)/E_Q$ tends to finite values as $r\to0$.  For $a>a_0$ the curves
$\frac{1}{A(r)}$ and $B(r)$ have one zero, and thus one horizon only,
at the same $r$, and then tend to infinity as $r\to0$, an analogous
behavior to the Schwarzschild black hole.  The function $E/E_Q$ tends
to finite values as $r\to0$.  See text for more details.}}
\label{K=2sqrt2gaussbonnetmodel}
\end{figure}

\begin{figure}
\centerline{\includegraphics[width=6.76in,height=4.95in]{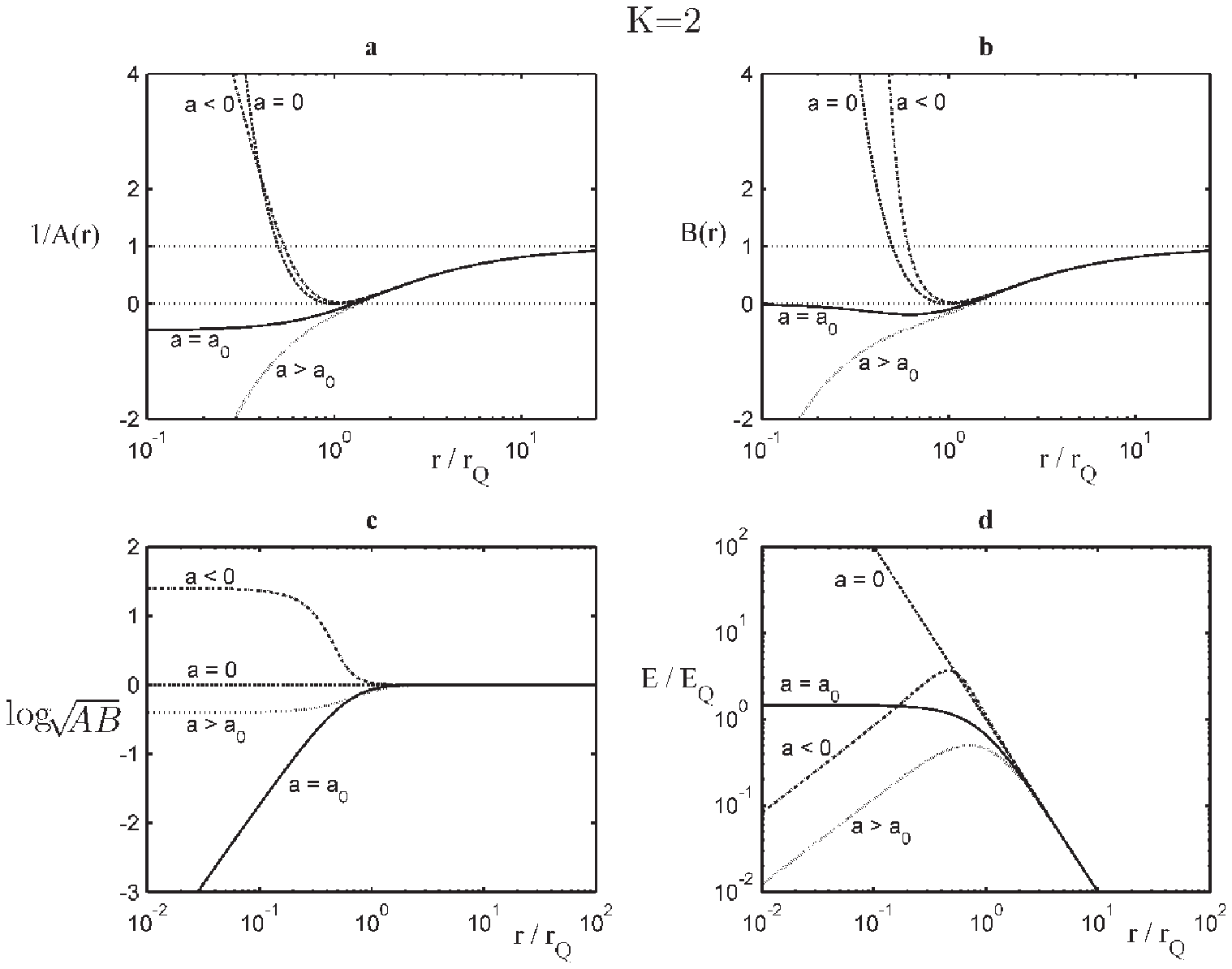}}
\caption {{\small Non-minimal solution of the Gauss-Bonnet
model, with $q_1=-q$, $q_2=2q$
and $q_3=-q$, of gravitational electrically charged objects
characterized by $a=2q/r_Q^2$, for $K=2$ --} {\small Plots (a), (b),
(c), and (d) depict the metric potentials $1/A(r)$ and $B(r)$, the
function $\log{\sqrt{A(r)B(r)}}$, and the electric field $E(r)/E_Q$,
respectively, as functions of $x=\frac{r}{r_Q}$, for solutions with
different values of the non-minimal quantity $a$.  The
extremal Reissner-Nordstr\"om black hole has $a=0$, the curves of
which are clearly shown in the plots. In this case the curves for
$1/A(r)$ and $B(r)$ have one double zero representing an extremal
horizon, and for $r\to\infty$ they go to one, while the curve
$E(r)/E_Q$ tends to zero, respectively, and when $r\to0$ these curves
tend to $-\infty$ (the electric field is in a logarithmic scale).  For
$a<0$ and $0<a<a_0$ the black holes behave quite similarly as the case
$a=0$, with one horizon, and the function $E(r)/E_Q$ tends to finite
values as $r\to0$. For $a=a_0$ the curve for $1/A(r)$ tends to a
finite negative value when $r \to 0$, and takes the value zero only
once. $B(r)$ has two zeros one at
the same point as $1/A(r)$, signaling there is only one horizon,
and the other at $r=0$, signaling the
presence of a singularity there. The function $E(r)/E_Q$ tends to
finite values as $r\to0$. For $a>a_0$ the curves $\frac{1}{A(r)}$ and
$B(r)$ have at the same $r$, one zero, and thus one horizon only, and
then tend to infinity as $r\to0$, an analogous behavior to the
Schwarzschild black hole. The function $E(r)/E_Q$ tends to finite
values as $r\to0$. See text for more details.}}
\label{K=2gaussbonnetmodel}
\end{figure}

\begin{figure}
\centerline{\includegraphics[width=6.76in,height=5.10in]{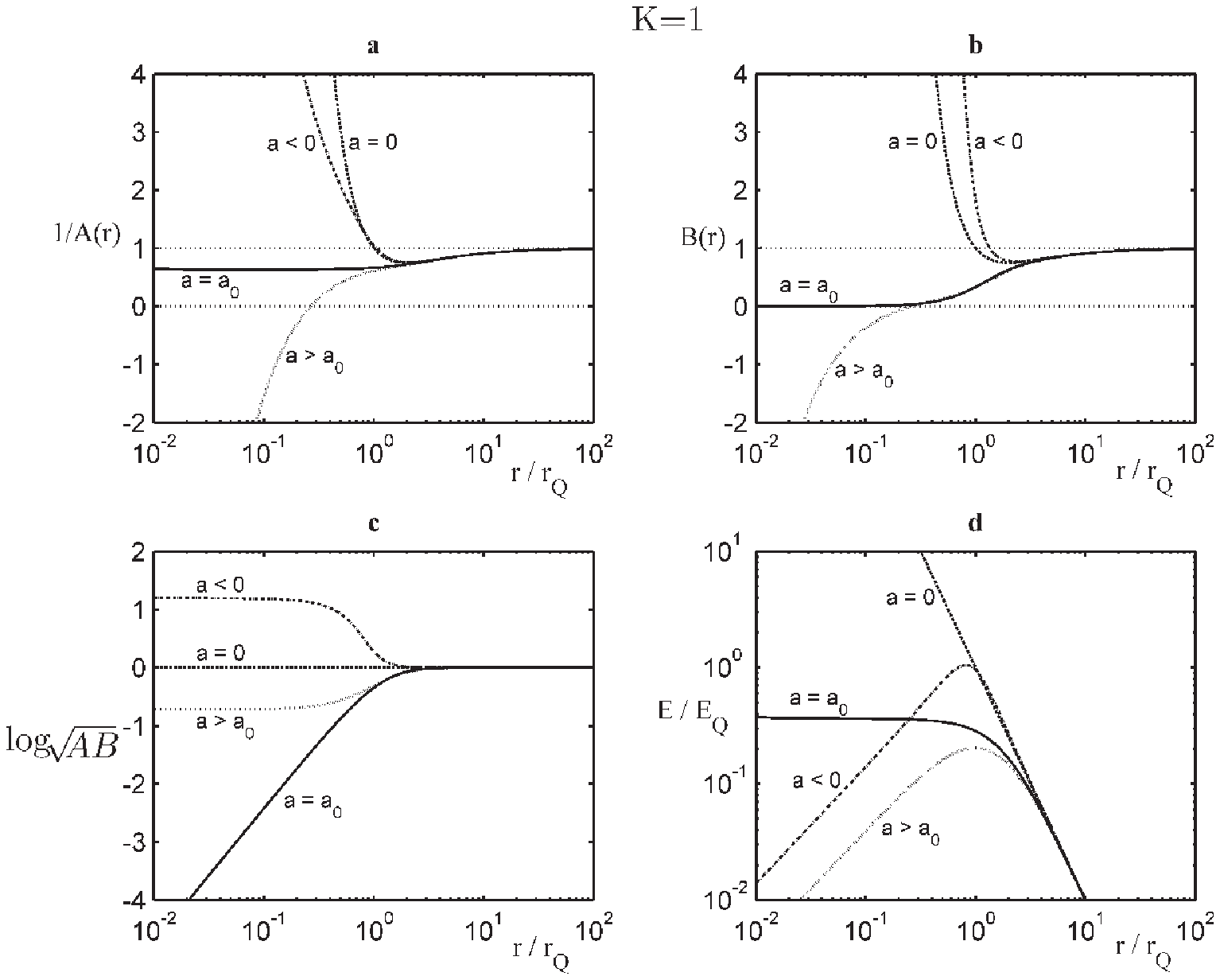}}
\caption[Figure5] {{\small
  Non-minimal solution of the Gauss-Bonnet
model, with $q_1=-q$, $q_2=2q$
and $q_3=-q$, of gravitational electrically charged objects
characterized by $a=2q/r_Q^2$, for $K=1$ ($K<2$) --}
{\small Plots (a), (b), (c), and (d) depict the metric potentials
$1/A(r)$ and $B(r)$, the function $\log{\sqrt{A(r)B(r)}}$, and the
electric field $E(r)/E_Q$, respectively, as functions of
$x=\frac{r}{r_Q}$, for solutions with different values of the
non-minimal quantity $a$.  The Reissner-Nordstr\"om naked
singularity has $a=0$, the curves of which are clearly shown in the
plots. In this case the curves for $1/A(r)$ and $B(r)$ have no zeros,
and for $r\to\infty$ they go to one, while the curve $E(r)/E_Q$ tends
to zero, respectively, and when $r\to0$ these curves tend to $-\infty$
(the electric field is in a logarithmic scale).  For $a<0$ and
$0<a<a_0$ the naked singularity behaves quite similarly as the case
$a=0$, and the function and $E(r)/E_Q$ tends to finite values as
$r\to0$. For $a=a_0$ the curve for $1/A(r)$ tends to a finite positive
value as $r \to 0$, while $B(r)$ has a zero at $r=0$, signaling the
presence of a singularity there. The function $E(r)/E_Q$ tends to
finite values as $r\to0$. For $a>a_0$ the curves $\frac{1}{A(r)}$ and
$B(r)$ have one zero at the same $r$, and thus one horizon only, and
then tend to infinity as $r\to0$, an analogous behavior to the
Schwarzschild black hole. Thus, by tuning the non-minimal quantity
$a$ one can turn a Reissner-Nordstr\"om naked singularity, which has
$a=0$, into a black hole, when $a>a_0$. The function $E(r)/E_Q$ tends
to finite values as $r\to0$. See text for more details.  }}
\label{K=1gaussbonnetmodel}
\end{figure}

\vskip 0.4cm
\noindent {(III) Power series expansion
with respect to $\frac{r_{Q}}{r}$}
\vskip 0.1cm
\noindent
The decomposition of the electric field yields
\begin{equation}
E(r) = \frac{Q}{r^2} \left[1 - a
\frac{r_M}{r_{Q}}\left(\frac{r_{Q}}{r}\right)^3
- \sum_{n=5}^{\infty} n b_n \left(\frac{r_Q}{r}\right)^{n-1} \right] \,,
\label{18}
\end{equation}
where the $b_n$ are defined below.
Infinity is a regular point for $N(r)$, thus, taking into
account (\ref{equaM2}) one obtains
the following decomposition of $N(r)$
\begin{equation}
N(r) = \frac{1}{A(r)} =  1 - \frac{{r_M}}{r} + \frac{r^2_Q}{r^2} -
a \frac{r_M}{4 r_Q}
\left( \frac{r_Q}{r} \right)^5 - \sum_{n=5}^{\infty}
b_n \left(\frac{r_Q}{r}\right)^{n+1} \,,
\label{17}
\end{equation}
where again, the $b_n$ are given below.
The function $\sigma(r)=\sqrt{A(r)B(r)}$ is equal to one in the
Schwarzschild and the Reissner-Nordstr{\"o}m cases, but not in
general.  When $q\neq0$ the logarithm of
this function can be represented by the
decomposition
\begin{eqnarray}
\ln \sigma(r) = - \frac{a}{4}\,\left(\frac{r_Q}{r}\right)^4 \left[ 1 {-}
\sum_{n=4}^{\infty} \frac{8nb_n}{n{+}3} \left(\frac{r_Q}{r}
\right)^{n{-}1}
\right. \nonumber\\
\left.
{+} \sum_{n=4}^{\infty} \sum_{m=4}^{\infty}\frac{4nm}{n{+}m{+}2}
b_n b_m \left(\frac{r_Q}{r}\right)^{n{+}m{-}2}
\right]\,.
\label{expansionofAB}
\end{eqnarray}
The $b_n$ coefficients can be taken from Eq. (\ref{17}), and
starting from $b_{5}$
can be found by the recurrence formula
\begin{equation}
b_{n{+}3} = - a \sum_{m=0}^{n{-}1}\left(\frac{n - m}{n + 3}\right)
b_{m} \ b_{n{-}m} \,,
\label{191}
\end{equation}
with
\begin{eqnarray}
b_0 = \frac{r_M}{r_Q} \,, \quad b_1 = -1 \,, \quad b_2 = b_3 = 0 \,,
\quad b_4 = a \frac{r_M}{4r_Q} \,, \quad b_5 = - \frac{a}{5} \,,
\quad\nonumber\\
b_6 = 0 \,, \quad b_7 = - a^2 \frac{r^2_M}{7 r^2_Q} \,, \quad  b_8 =
a^2 \frac{9 r_M}{32r_Q} \,, ...\,.
\label{19}
\end{eqnarray}
The decompositions (\ref{18})-(\ref{expansionofAB}) are regular at
$r=\infty$ and absolutely converge in the interval $r > r_Q\, H(a)$,
where $H(a) \equiv \lim_{n \to \infty} |\frac{b_{n+1}}{b_n}| $. Note
that the terms $b_0$ and $b_1$ are the Schwarzschild and
Reissner-Nordstr{\"o}m terms, respectively, and that the $b_n$ for
$n\geq2$ are the terms that give the post--Reissner-Nordstr{\"o}m
behavior. Note also that for $q=0$, i.e., the Reissner-Nordstr{\"o}m
case (or the Schwarzschild case when, further, $Q=0$), the function
$\sigma$ in (\ref{expansionofAB}) is equal to one, as it should.  For
$r\to\infty$, Eqs. (\ref{18})-(\ref{expansionofAB}) give useful
asymptotic formulas, showing that, for $E(r)$ and $N(r)$, the first
post-Reissner-Nordstr\"om terms are of fifth order in
$\left(\frac{r_Q}{r} \right)$, and that the decomposition for
$\log\sigma$ starts with a term of fourth order.  When
necessary one should
convert from $N$ and $\sigma$ to $A$ and $B$.  Numerical calculations,
see Figs. \ref{K=2sqrt2gaussbonnetmodel}-\ref{K=1gaussbonnetmodel},
confirm that the corresponding curves tend to the
corresponding horizontal asymptotes, when $r$ goes to infinity,
i.e., $E(r)$
tends to zero, and $\frac{1}{A(r)}$ and $B(r)$ tend to 1.  It follows
from (\ref{18}) that, for positive $q$, $q>0$, the curvature coupling
effects on the electric field are analogous to a dielectric medium,
since the asymptotic electric field effectively decreases. For
negative $q$, $q<0$, there are no solutions with $N(r)$ regular at
$r=0$, and, since we are mostly interested
in regular or quasi-regular solutions, we do not fully discuss this case.

\subsubsection{The dynamical system associated with the model}
We now study this model, specified through Eq. (\ref{modeladopted}),
using a dynamical system analysis.

\vskip 0.5cm
\noindent{(I) First analysis: The plots and numerics}

\vskip 0.3cm

\noindent{\it (A) Key dynamic equation:}
In order to find the regular solutions $E(r)$, $N(r)$ and $\sigma(r)$
in the whole interval
$0<r<\infty$ let us transform the master Eqs. (\ref{11e1}),
(\ref{12e11}) and (\ref{13e1}) to the
independent variable $x \equiv \frac{r}{r_{Q}}$, a dimensionless radius,
and to the unknown dimensionless
function $y(x)$ given in Eqs. (\ref{NRNnew}) and (\ref{yRN}), i.e.,
\begin{equation}
y(x) \equiv x [1-N(x)]\,.
\label{ydef}
\end{equation}
The physical interpretation of $y(x)$ is
connected with the so-called effective mass of the object, $M(r)$,
see Eq. (\ref{NYXRN}).
Putting these definitions into Eq. (\ref{13e1}), we
obtain the following key equation
\begin{equation}
\frac{d y(x)}{d x} = \frac{x}{x^3 + a y(x)} \,. \label{key}
\end{equation}
This equation is indeed a key one, since using its solution, $y(x)$,
we can represent explicitly the electric field by
\begin{equation}
E(x) =  E_Q \, Z(x) \,,\;\; {\rm with}\; \quad Z(x)
\equiv \frac{1}{x^2 + a y(x)/x}\,,
\label{newE}
\end{equation}
the metric function $N(x)$ by Eq. (\ref{NYXRN}), and $\sigma(x)$
by the integral form
\begin{equation}
\ln{\sigma(x)} = \ln \sqrt{A(x)\,B(x)} = a \int^x_{\infty}
\frac{d x'}{x'}\,Z^2(x') \,. \label{newAB}
\end{equation}
Moreover, $A$ follows from $A=1/N$ and $B$ from $B=\sigma^2/A$, i.e.,
\begin{equation}
B(x) = \left[1-\frac{y(x)}{x} \right] \exp{\left\{2a \int^x_{\infty}
\frac{d
x'}{x'}\,Z^2(x')\right\}} \,. \label{newB}
\end{equation}
Note that the function $y(x)$ is also a function of the quantity $a$,
so in general should be written as $y(x,a)$.

\begin{figure}
\centerline{\includegraphics[width=6.76in,height=4.72in]{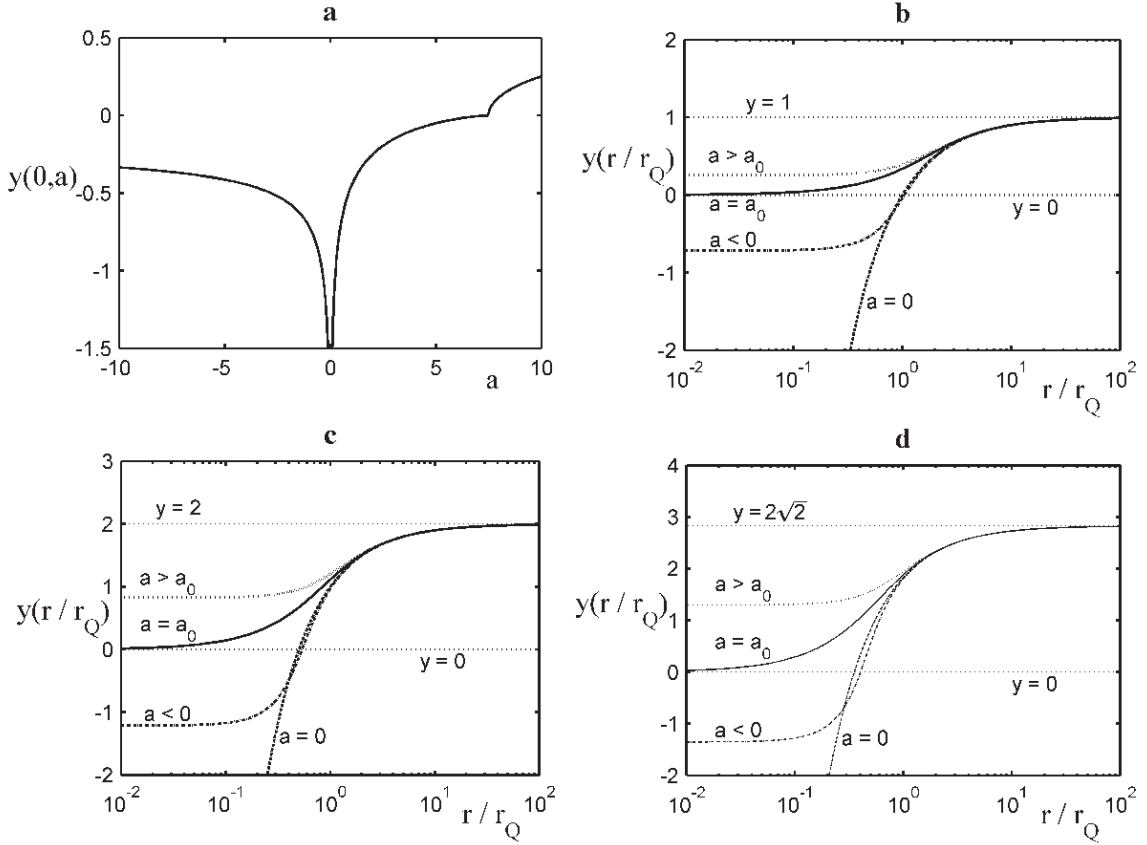}}
\caption{{\small Reduced mass profiles $y(0,a)$ and $y(x,a)$ for the
non-minimal Gauss-Bonnet model, with $q_1=-q$, $q_2=2q$ and $q_3=-q$,
of gravitational electrically charged objects
characterized by $a=2q/r_Q^2$ --} {\small
Plots of the curves $y(0,a)$ and $y(x,a)$
are shown, with $x=r/r_Q$.
For $a_0<a$ (with $a_0>0$) the characteristic curve $y = - x^3/a$ is not
intercepted by the function $y(x,a)$.
For
$a=a_0$ the intersection takes place at $x=0$. For $0<a<a_0$ the point
of crossing floats along the characteristic curve, and the integral
curve has two branches. Finally, for $a<0$ there is no intersection of
$y(x,a)$ with the characteristic curve. In the plot $y(0,a)$ it is
shown explicitly the existence of a point obeying $y(0,a_0)=0$ with
positive $a_0$. }}
\label{yingausbonnetmodel}
\end{figure}

\vskip 0.5cm
\noindent{\it (B) Three typical cases:}
In this problem there are two independent dimensionless quantities,
constructed from $r_M$, $r_Q$, and $r_q$, namely $a$ and $K$, see
Eqs. (\ref{aratio}) and (\ref{K0}).  Note that $K$, besides
fixing the ratio between the total mass and the charge of the object,
gives the value of the dimensionless mass $y(x)$ at $r=\infty$, since
$y(\infty)=K$.  Taking into account the quantity $K$, and in
conformity with the Reissner-Nordstr\"om solution, let us distinguish
three different situations, (i) $K>2$, (ii) $K=2$, (iii) $K<2$, within
each situation the quantity $a$ can vary from zero to infinity.
Figs. \ref{K=2sqrt2gaussbonnetmodel}-\ref{K=1gaussbonnetmodel} display
typical cases in each situation, and Fig.  \ref{yingausbonnetmodel}
shows the behavior of $y(x)$. In slightly more detail:
(i) $K>2$ (i.e., $r_M>2r_Q$): For $K>2$, we use $K= 2 \sqrt{2}$
as a typical value for the numerical analysis, see the plots in
Fig. \ref{K=2sqrt2gaussbonnetmodel},
(see Fig. \ref{K=2sqrt2gaussbonnetmodel}a
for $\frac{1}{A(r)}$, Fig. \ref{K=2sqrt2gaussbonnetmodel}b for $B(r)$,
Fig. \ref{K=2sqrt2gaussbonnetmodel}c for $\ln \sqrt{A(r)B(r)}$,
and Fig. \ref{K=2sqrt2gaussbonnetmodel}d for $E(r)$). When
$a=0$ this case gives the usual Reissner-Nordstr\"om black hole with
two horizons. For other $a$s there are also black holes, some with
different properties.
(ii) $K=2$ (i.e., $r_M = 2 r_Q$): For $K= 2$ see the plots in
Fig. \ref{K=2gaussbonnetmodel},
(see Fig. \ref{K=2gaussbonnetmodel}a
for $\frac{1}{A(r)}$, Fig. \ref{K=2gaussbonnetmodel}b for $B(r)$,
Fig. \ref{K=2gaussbonnetmodel}c for $\ln \sqrt{A(r)B(r)}$,
and Fig. \ref{K=2gaussbonnetmodel}d for $E(r)$). When $a=0$
this case gives the extreme Reissner-Nordstr\"om black hole with one
horizon. For other $a$s there are also interesting solutions with
black holes.
(iii) $K<2$ (i.e., $r_M < 2 r_Q$): For $K<2$, we use $K=1$
as a typical value for the numerical analysis, see the plots in
Fig. \ref{K=1gaussbonnetmodel},
(see Fig. \ref{K=1gaussbonnetmodel}a for $\frac{1}{A(r)}$,
Fig. \ref{K=1gaussbonnetmodel}b for
$B(r)$, Fig. \ref{K=1gaussbonnetmodel}c for $\ln \sqrt{A(r)B(r)}$,
and Fig. \ref{K=1gaussbonnetmodel}d for $E(r)$). When $a=0$
this case gives a  Reissner-Nordstr\"om naked singularity, a
solution without horizons. For other $a$s there are also
solutions.

\vskip 0.5cm
\noindent{\it (C) Scaling of the key equation:}
The key Eq. (\ref{key}) remains invariant after the following
scale transformations:
\begin{equation}
x \to \frac{1}{K}\, x\;,\quad y \to K \,y\;,\quad a \to
\frac{1}{K^4} \,a\;. \label{scaletransformations}
\end{equation}
Thus, the critical values of $a$ that one may eventually encounter
when $K=1$, are also critical values that one can easily find for
arbitrary $K$ using the formula $a_0(K)=\frac{1}{K^4}\,a_0(1)$.

\vskip 0.5cm
\noindent{(II) Second analysis: Critical properties
of the family of the solutions}
\vskip 0.3cm

\noindent{\it (A) About the mass function $y(x,a)$
when $x=0$, $y(0,a)$ - the critical value of the quantity $a$, $a_0$:}
We now write explicitly that $y$ is a function of both $x$ and $a$,
$y=y(x,a)$ since this is important to our analysis.  Plots of this
dimensionless mass function $y(x,a)$ are displayed in Fig.
\ref{yingausbonnetmodel}. In
Fig. \ref{yingausbonnetmodel}a, $y(x,a)$ is shown for several values of
$a$,
and in Fig.
\ref{yingausbonnetmodel}b, a plot for $y(0,a)$ as a function of $a$ is
shown.
A simple
qualitative analysis shows that the mass function $y(x,a)$ at the
central point $x=0$, i.e. $y(0,a)$, as a function of the quantity $a$,
has to possess a zero. Indeed, when $a=0$, $y(x,0) = K - \frac{1}{x}$,
(where, recall, $K \equiv \frac{r_M}{r_Q}$), corresponding thus to the
Reissner-Nordstr\"om solution.  When $a = \infty$, $y(x, \infty) = K$,
corresponding thus to the Schwarzschild solution.  Note as well that
when $a = -\infty$, $y(x,-\infty) = {\rm constant} $. In addition,
when $x= \infty$, $y(\infty, a) = K$, a condition at infinity that
holds for arbitrary $a$. In order to prove our assertion, that
$y(0,a)$ as a function of $a$ possesses a zero, consider then $y(0,a)$
as a function of the quantity $a$ in the interval $0<a <\infty$.  One
can see, that $y(0,0) = - \infty$, and $y(0, \infty) = K > 0$.
Supposing that $y(0,a)$ is continuous in such an interval, one can
conclude that there exists at least one specific value of the $a$
quantity for which $y(0,a_0)= 0$, where $a_0$ is the value of $a$ for
which $y(0,a_0)= 0$.  Fig. \ref{yingausbonnetmodel}b
shows that, for $K=1$ the zero of the
function $y(0,a)=0$ happens, when $a \equiv a_0 \simeq 7.49$.  But the
most interesting fact is that the curve $y=y(0,a)$ displays a
discontinuity in the first derivative with respect to $a$ just at
$a=a_0$. One can see explicitly a finite jump of the derivative at
this point, $a=a_0$.  Nevertheless, the function $y= y(0,a)$ itself is
continuous at this point. That is why $a_0$ is a critical value of
the quantity $a$. For other $K$s the critical values can be found
using the scaling properties (\ref{scaletransformations}), yielding
$a_0(K) = \frac{7.49}{K^4}$. For instance $a_0(2) = 0.468$ and, for
the typical case we study, $a_0(2\sqrt{2})= 0.117$.  For negative $a$,
$y(0,a)\to0$ asymptotically when $a\to-\infty$.

\vskip 0.5cm
\noindent{\it (B) The critical points of the associated autonomous
two-dimensional dynamical system:}
The key Eq.  (\ref{key}) can be put as an autonomous dynamical system
\begin{equation}
\dot{y} = x \,, \quad \dot{x} = x^3 + a\,y \,,
\label{autonomoussystem}
\end{equation}
where $\dot{}\equiv \frac{d\,}{d\tau}$ and $\tau$ is an auxiliary
parameter. In Eq. (\ref{autonomoussystem}) there is one
critical point at
\begin{equation}
(x,y)=(0,0)\,. \label{criticalpointsy}
\end{equation}
In the vicinity of this critical point the variables $x$ and $y$
are connected by the relation $x^2 - a y^2 = {\rm constant}$,
which means this point is a saddle point when $a$ is positive, and
a center when $a$ is negative. If $a>0$ there are two separatrices
$y = \pm \frac{x}{\sqrt{a}}$.
The equation for $N(r)=\frac{1}{A(r)}$ can also be written as
dynamical,
\begin{equation}
\dot{N} =  \left(1 - N \right)
r^2 + 2q \left(1 - N \right)^2 - r^2_Q \,, \quad \dot{r}
= r \left[ r^2 + 2q \left(1 - N \right)\right] \,,
\label{21}
\end{equation}
which is much more complicated than Eq. (\ref{autonomoussystem})
for $y(x)$.  Nevertheless, if $q$ is positive (i.e., $a$ is positive),
one can find the critical points immediately,
\begin{equation}
\left(r, \ N \right)=\left(0, \ 1 \pm \frac{r_Q}{r_q}
\right)\,. \label{criticalpointsy1/A}
\end{equation}
In order to present the integral curves for the total interval of the
auxiliary parameter $\tau$, we resort to numerical calculations. The
results are presented in
Figs. \ref{K=2sqrt2gaussbonnetmodel}-\ref{K=1gaussbonnetmodel}.
It is clear that for the
critical $a=a_0$ the curve for $N(r) = \frac{1}{A(r)}$ tends to one
when $r \to \infty$, and takes a finite value $\frac{1}{A(0)} = 1 -
\frac{1}{\sqrt{a_0}}$ at the center of the object, $r=0$. This
critical curve is a separatrix between the curves having $a>a_0$ and
those having $a<a_0$. The same type of behavior happens with the curves
for $B(r)$. One also has that for $a = a_0$ there exists a unique
integral curve for $y(x)$ with asymptotic value given by $y(\infty) =
\frac{r_M}{r_Q}$, and for $r\to0$ one has $y(0,a_0)=0$. Again, this
curve behaves as a separatrix. Similar reasoning goes to the curve
$E(r)$.  In other words, the critical point $N = 1 - \frac{r_Q}{r_q}$,
given in Eq. (\ref{criticalpointsy1/A}), is just a saddle point,
corresponding to the critical value $a_0$, obtained numerically, and
it is the final point of the unique integral curve, which coincides
with the separatrix of the function $y = +
\frac{x}{\sqrt{a_0}}$ at small $x$.  When $q<0$ and, thus, $a$ is
negative, there are no regular or quasi-regular
solutions to the master equations
for the whole interval $0\leq r < \infty$.

\vskip 0.5cm
\noindent{\it (C) Vertical asymptotes and the electric barrier:}
From Eq. (\ref{key}) one sees that, when $y(x,a)=-\frac{x^3}{a}$,
the derivative $y^{\prime}(x,a)$ becomes infinite and vertical
asymptotes appear in the graph of $y(x)$ versus $x$. At the point
$x=x^{*}$, for which $y(x^{*},a)=-\frac{x^{*3}}{a}$, the electric
field (\ref{newE}) becomes infinite, and so these vertical asymptotes
can be interpreted in terms of an electric barrier.

When $a<0$,
Figs. \ref{K=2sqrt2gaussbonnetmodel}d,
\ref{K=2gaussbonnetmodel}d, and
\ref{K=1gaussbonnetmodel}d show that for all three values of
$K$ the curve for the electric field has a form of a finite
barrier. Indeed for $r\to\infty$ the electric field $E(r)$ tends to
zero, then at some radius it reaches a maximum, the barrier height,
and finally goes to a finite positive value when $r\to0$. A charged
test particle with sufficiently high energy can overcome this barrier,
be trapped in the potential well and then oscillate inside. When
$a=0$, more specifically, when $a\to0_-$, the barrier height increases
and tends towards the center $r\to0$. For $a=0$ (see again Figs.
\ref{K=2sqrt2gaussbonnetmodel}d,
\ref{K=2gaussbonnetmodel}d, and
\ref{K=1gaussbonnetmodel}d)
one obtains the standard, Reissner-Nordstr\"om, behavior,
$E(r)=\frac{Q}{r^2}$. This means that the electric barrier has become
infinite, taking its maximum value (an infinite value) at the center
$r=0$. In other words, the vertical asymptote for the electric field
appears at $r=0$.  Finally, when $a>0$ this vertical asymptote and the
position of the infinite electric barrier shift from the center
towards positive $r$ values, then stop at some value of the quantity
$a$, drift back again to smaller values of $r$ and, finally, the
infinite electric barrier disappears at $a\geq a_0$,
(with $a_0>0$). When $a=a_0$
the curve tends to the horizontal asymptote at $r \to 0$, there is yet
no trap. When $a > a_0$, one can see a finite electric barrier with
the corresponding traps, for all values of $K$.

Since the integrals in (\ref{newAB}) and (\ref{newB}) diverge when
$E(r)$ is discontinuous, then in searching for solutions with regular
functions $E(r)$, $\frac{1}{A(r)}$, and $B(r)$, we should reject all
the cases where vertical asymptotes appear.  For $a \geq a_0$,
numerical calculations show that vertical asymptotes do not appear.
So we will mainly consider solutions for this range of the quantity
$a$, i.e., $a\geq a_0$.

\vskip 0.5cm
\noindent{\it (D) Horizons:} To analyze the
$a\geq a_0$ case, with a non-singular electric field
and possible non-singular metric potentials
$1/A$ and $B$, we have studied previously, for comparison,
the usual static spacetimes with $a=0$, i.e., the Schwarzschild and
Reissner-Nordstr\"om spacetimes, which are special solutions of Eq.
(\ref{metric1}). One has for these that
$A(r)\,B(r)=1$.
For $a\geq a_0$ one
sees from (\ref{expansionofAB}) that $A(r)\,B(r) \neq 1$,
in contrast to the $a=0$ case. Nevertheless, as in the $a=0$ case,
horizons are still given by the condition
$\frac{1}{A(r)} = 0$, or $y(x)=x$. We have analyzed this
numerically. For $a>a_0$ the results are the following:
(i) $K>2$ (i.e., $r_M>2r_Q$)):
A typical case is $K= 2\sqrt{2}$, see
Fig. \ref{K=2sqrt2gaussbonnetmodel}.
One finds that for $a>a_0 = 0.117$, there is only one horizon.
(ii) $K=2$ (i.e., $r_M = 2 r_Q$):
For $K= 2$, see Fig. \ref{K=2gaussbonnetmodel}.
One finds that for $a>a_0=0.468$,
there is one horizon also.
(iii) $K<2$ (i.e., $r_M < 2 r_Q$)):
A typical case is $K= 1$, see Fig. \ref{K=1gaussbonnetmodel}.
One finds that for $a>a_0 = 7.49$, there is only one horizon also.

Note that
Figs. \ref{K=2sqrt2gaussbonnetmodel}-\ref{K=1gaussbonnetmodel} show
that when $a>a_0$, for all formal possibilities ($K>2$, $K=2$, $K<2$)
the curves $1/A$ tend monotonically to minus infinity and cross the
line $1/A =0$ only once. Thus, for arbitrary $a>a_0$ the plots of
$1/A$ are continuous, irregular at the center and characterized by one
horizon. These solutions have thus an analogous behavior to the
Schwarzschild black hole. Moreover, by tuning the non-minimal quantity
$a$ one can turn a Reissner-Nordstr\"om naked singularity, with $K<2$
and $a=0$, into a black hole, when $a>a_0$.  Now, since the $a=a_0$ is
a very special case, we discuss it in particular.

\vskip 0.5cm
\noindent{\it (E) The solution with $a=a_0$:}
In the framework of the model under discussion, i.e., when
$q_1=-q$, $q_2=2q$, and $q_3=-q$, all the three functions, $E(r)$,
$\frac{1}{A(r)}$, and $B(r)$, are regular in the interval
$0 \leq r <\infty$ if and only if
$a=a_0(K)=\frac{7.49}{K^4}$ (recall
$a\equiv\frac{2q}{r^2_Q}$).   This means that we
deal with a one-parameter family of exact solutions, the arbitrary
quantity being $K=\frac{r_M}{r_Q}$, and all the curvature coupling
constants, $q_1$, $q_2$ and $q_3$, being expressed explicitly
via $K$. The critical $q$ corresponding to $a_0$ is thus, with the
help of Eqs. (\ref{rq})-(\ref{K0}), given by,
$q_{a_0}=7.49 Q^6 / 32 GM^4$.
We now consider these solutions in more detail:
(i) There are two different solutions, corresponding to the
separatrices $y(x)= \pm \frac{x}{\sqrt{a_0}}$ at small $x$. The
physical solution, the solution that gives the appropriate limit when
$x \to \infty$ and has no jumps on the derivative of the characteristic
curve $ay(x)=-x^3$, is $y(x)=+\frac{x}{\sqrt{a_0}}$. The plots of
$\frac{1}{A(r)}$, $B(r)$ and $E(r)$ are displayed in
Figs. \ref{K=2sqrt2gaussbonnetmodel}-\ref{K=1gaussbonnetmodel}.
(ii) The solutions, which we discuss, are characterized by
finite values at the center, and $E(0) \neq 0$, $A(0) \neq 0$, and $N(0)
\neq 0$, but $\sigma(0) =0$ and $B(0)=0$.
(iii) For this model $\frac{1}{A(r)} \neq B(r)$, and there are
two distinct critical radii; first, the radius for which
$\frac{1}{A(r)}=N(r)=0$; second, the
radius for which $B(r)=0$ which signals an infinite redshift surface,
and an event horizon in the case of static spacetimes,
such as the ones we are treating here.
(iv) For $\frac{1}{A(r)}=0$, one finds that such a solution
exists when $r_q < r_Q$, i.e., $a_0 <1$; this is a necessary condition
(see, e.g., (\ref{small2})). Moreover, within this case it is
possible to have such a zero when $K^4 > 7.49$, i.e., $K>1.65$ or $r_M >
1.65 r_Q$. For $r_q > r_Q$, one has, $\frac{1}{A(0)} = 1 -
\frac{r_Q}{r_q}$ is positive and without zeros.
(v) For the infinite redshift surface and event horizon,
$B(r)=0$, one finds that when
$a=a_0$, for the critical quantity, the function $B(r)$ is zero both
at $r=0$ and at the radius for which $\frac{1}{A(r)}=0$.
(vi) The Kretschmann scalar diverges at $r=0$, so although
the metric functions are regular, spacetime is not, time stops.

\vskip 0.5cm
\noindent{\it (F) Remark:}
Some important aspects derived from qualitative and numerical
analyses of this non-minimal model, with $q_1=-q$, $q_2=2q$
and $q_3=-q$, can be found in \cite{MHS}.
Our results are in thorough concordance with this initial analysis.
We have supplemented those aspects on several grounds, of which
we stress briefly three novel details obtained here:
\noindent (i) We have found a complete converging decomposition
of the solution of the Abel equation based on the recurrence
formula (\ref{191}). This gives us not only the asymptotic
decompositions for $r \to \infty$, but also the possibility to link
the limiting formulas for $r\to0$ and $r \to \infty$
(see (\ref{18})-(\ref{expansionofAB})).
\noindent (ii) We have formulated and discussed the problem of
the infinite electric barrier, associated with the vertical asymptote
appearing when $x^3+ay(x,a)=0$, thus completing physically and
mathematically the analysis given in \cite{MHS}.
\noindent (iii) We have found, first, the critical value of the
non-minimal parameter $q$, namely, $q_{a_0}=7.49 Q^6 / 32 GM^4$,
second, the scaling law of the critical parameter $q_{a_0}$ for different
values of the asymptotic quantity $K=r_{M}/r_{Q}$, namely,
$q_{a_0}(K)=7.49 / K^4$, third, the significance of the choice for $K$,
namely, $K>2$, $K=2$, and $K<2$, in the qualitative analysis.

\subsection{The integrable model: $q_1\equiv -q$, $q_2=q$, $q_3=0$
(i.e., $q_1+q_2+q_3=0$ and $q_3=0$)}
\label{nonamemodel}

\subsubsection{Preliminaries}

Now, we consider the model with
\begin{equation}
q_1\equiv -q, \,q_2=q, \,q_3=0\;\;
({\rm i.e.}, \,q_1+q_2+q_3=0,\,q_3=0)\,,
\end{equation}
which, as we have seen, has the property that $E(r)$ decouples from
$N(r)$ and we deal with a cubic equation for the determination of the
electric field.  Since the basic feature of the model is that it is
integrable, we call it the integrable model.  For such a model the
susceptibility tensor $\chi^{ikmn}$ can be written in terms of the
Einstein tensor $G^{ik} \equiv R^{ik}-\frac{1}{2}R g^{ik}$ as follows,
$
\chi^{ikmn} = \frac{q}{2}\left[
G^{im} g^{kn} - G^{in} g^{km} + G^{kn} g^{im} - G^{km} g^{in} \right],
$
where we have put $q_1=-q_2\equiv -q$. Thus, the model becomes
one-parametric and we can introduce the dimensionfull
quantities defined above,
$r_Q$, $E_Q$, $r_q$, and the dimensionless quantities
$a$ and $K$. We assume that $E_Q$ inherits
the sign of the charge $Q$ and the quantity $a$ can be positive or
negative depending on the sign of $q$.  Then, we introduce the two
dimensionless variables, the normalized radius $x$ and the normalized
electric field $Z(x)$, defined in (\ref{dimssRN}).
In terms of these, the system of key equations can be rewritten as
\begin{equation}
a (x^2 +a) Z^3 - 2a Z^2 - x^2 Z + 1 = 0 \,,
\label{1eqdim}
\end{equation}
\begin{equation}
x N^{\prime}(x) + N \left[ 1 - a x (Z^2)^{\prime}(x)
\right] = 1 -  2 Z + (x^2+a) Z^2 \,,
\label{2eqdim}
\end{equation}
\begin{equation}
\frac{\sigma^{\prime}(x)}{\sigma} = - a (Z^2)^{\prime}(x) \,.
\label{3eqdim}
\end{equation}
Clearly, Eq. (\ref{1eqdim}) for $Z(x)$ is the
key equation for finding $N(x)$ and $\sigma(x)$ from
(\ref{2eqdim}) and (\ref{3eqdim}), respectively.
If instead of $N(x)$ one uses $y(x)$ then
Eq. (\ref{2eqdim}) gives an equation for $\frac{dy}{dx}$ of the type
of Eq. (\ref{ydashRN}) or Eq. (\ref{key}), but more complicated,
which for this analysis is not very illuminating.
For this model it is better to start analyzing the electric field
$E(r)$, or its redefinition $Z(x)$.

\subsubsection{Electric field}

\begin{figure}
\centerline{\includegraphics[width=6.76in,height=6.75in]{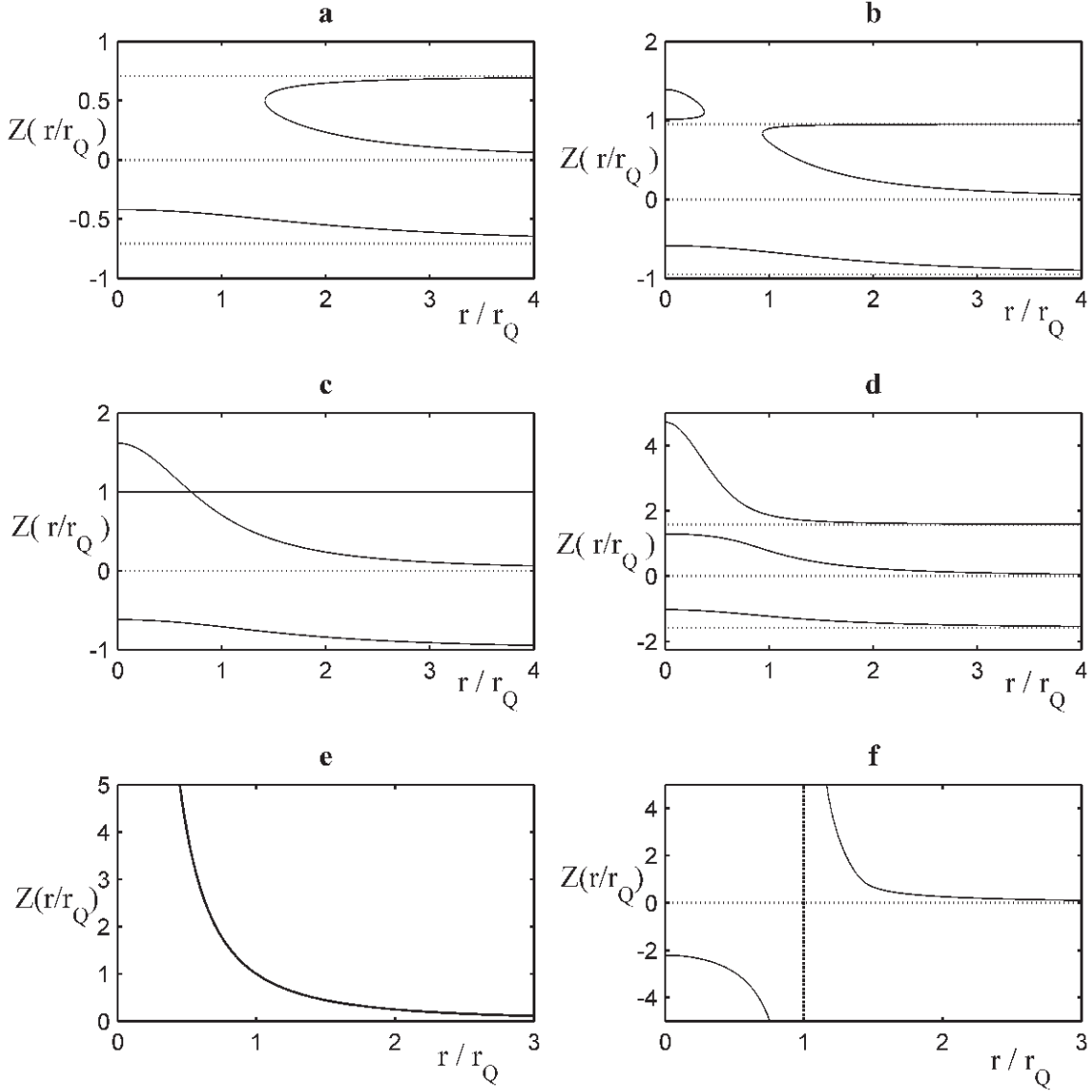}}
\caption {{\small It is shown, in the non-minimal integrable model, with
$q_1=-q_2\equiv -q$ and $q_3=0$, the rescaled electric field
$Z(x,a)$, $(Z(x,a)\equiv E(r)/E_Q)$, as a function of the scaled
radius $x$ $(x\equiv r/r_Q)$ of gravitational electrically charged
objects characterized by $a=\frac{2q}{r^2_Q}$ -- }
{\small   (a) Displays the solution $Z(x,a)$ when the
non-minimal quantity $a$ satisfies the inequality
$ \frac{32}{27} <a < \infty$. There are three real branches of the
solution $Z(x,a)$. Nevertheless, only one of them is defined on
the whole interval $0 \leq x < \infty$. Only one branch tends to
the horizontal asymptote $Z=0$ at $x \to \infty$.   (b)
Displays the solution $Z(x,a)$ in the case $1<a \leq \frac{32}{27}$.
There are three real starting points $Z_1(0,a)$,
$Z_2(0,a)$, $Z_3(0,a)$ for the three corresponding branches of the
solution $Z(x,a)$. Nevertheless, two branches are not continuous,
only the third being defined on the whole interval $ 0 \leq x <
\infty$.   (c) Displays the solution $Z(x,a)$ for the important case
$a\equiv a_0=1$, the Fibonacci soliton.
(d) Displays the solution $Z(x,a)$ in the case $ 0
< a < 1$.   (e) Displays the solution $Z(x,a)$ in the case
$a=0$ which is a Coulombian electric field. The curve is not
continuous.   (f) Gives an example of the solution for negative
$a$ with a vertical asymptote.  When $a$ tends to zero remaining
negative, the vertical asymptote shifts towards the line $x=0$,
and the solution $Z(x,a \to 0_{-})$ converts, finally, into the
Coulombian solution. At large values of $x$ the plots of the function
$Z(x,a<0)$ tend to the Coulombian curve $Z(x,0)$ for all values of
the quantity $a$. }} \label{electricfieldinnonamemodel}
\end{figure}

Consider now Eq. (\ref{1eqdim}) in detail.  Eq.
(\ref{1eqdim}) is a one-parameter algebraic equation of third order
for the dimensionless electric field. Below we denote its solution as
$Z(x,a)$.  The function $Z(x,a)$ can be generally presented by the
well-known Cardano formula, nevertheless we prefer to analyze
qualitatively this one-parameter family of solutions. Depending on the
value of the quantity $a$ the solution $Z(x,a)$ can possess one or
three real branches.  The corresponding plots are presented in
Fig. \ref{electricfieldinnonamemodel}.  When $a=0$, one obtains, as it
should, the Coulombian solution $Z(x,0)=\frac{1}{x^2}$.  The curves
displaying $Z(x,a)$ for non-vanishing values of the quantity $a$ are
more sophisticated.

\vskip 0.2cm
\noindent
{\it (i) $a>0$: }
When $a$ is positive, the functions $Z(x,a)$ take finite values
for all values of the quantity $a$, see the curves {\rm a},
{\rm b}, {\rm c}, {\rm d} in
Fig. \ref{electricfieldinnonamemodel}.  The initial values
$Z(0,a)$ satisfy the cubic equation $ a^2 Z^3(0,a) - 2a Z^2(0,a) +
1=0$. There is only one real solution of this cubic equation for
$a>0$, if the discriminant ${\cal D}=\frac{1}{108a^5}(27a-32)$ is
positive, i.e., when $ \frac{32}{27} < a < \infty$, see box
${\rm a}$ of Fig. \ref{electricfieldinnonamemodel}
for details.  This plot displays three real
branches of the solution $Z(x,a)$, nevertheless, only one of them is
defined on the whole interval $0 \leq x < \infty$. Two other branches
are defined for $x \geq x_{{\rm min}}(a)$ only and contact at the
point $x=x_{{\rm min}}(a)$. Only one branch tends to the horizontal
asymptote $Z=0$ at $x \to \infty$.
When $0 < a \leq \frac{32}{27}$, the discriminant ${\cal D}$ is
negative or equal to zero, which guarantees that there are three real
starting points, $Z_1(0,a)$, $Z_2(0,a)$, $Z_3(0,a)$ for the three
corresponding branches of the solution $Z(x,a)$, see the curves in
boxes {\rm b}, {\rm c}, {\rm d} of Fig.
\ref{electricfieldinnonamemodel}.
Nevertheless, when $ 1< a \leq \frac{32}{27} $,
two branches of the solution $Z(x,a)$ are not continuous, only the
third being defined on the whole interval $ 0 \leq x < \infty$, see
box {\rm b} of Fig. \ref{electricfieldinnonamemodel}.
When $ 0 < a \leq 1 $ all three
branches are continuous and are defined on the whole interval $ 0 \leq
x < \infty$, one of them is asymptotically Coulombian.  There are
three horizontal lines $Z= - \frac{1}{\sqrt{|a|}}$, $Z=0$ and $Z=
+\frac{1}{\sqrt{|a|}}$, which yield distinct
ranges for the
functions $Z_1(x,a)$, $Z_2(x,a)$, and $Z_3(x,a)$.  Clearly, the curve of
Coulombian type is in-between the separatrices $Z=0$ and $Z= +
\frac{1}{\sqrt{|a|}}$.
The model with critical $a$, call it $a_0$ again, is the one that has 
$a=1$,
i.e., $a\equiv a_0=1$.This model can be solved
analytically, and we consider this case in detail below.
Finally, when $a$ tends to zero remaining positive, the starting
points $Z_1(0,a)$ tend to minus infinity, and $Z_2(0,a)$, $Z_3(0,a)$
grow infinitely. Clearly, at $a \to 0_{+}$ the branch $Z_2(x,a)$
is the only branch that remains visible at finite values of $x$,
and is Coulombian.

\vskip 0.2cm
\noindent
{\it (ii) $a=0$:}
When $a=0$, one obtains the Coulombian solution
$Z(x,0)=\frac{1}{x^2}$,
see the curve in box ${\rm e}$ of Fig.
\ref{electricfieldinnonamemodel}.

\vskip 0.2cm
\noindent
{\it (iii) $a<0$:}
When $a$ is negative, the coefficient $(x^2+a)$ in the first term of
Eq. (\ref{1eqdim}) vanishes at $x = \sqrt{|a|}$ and the line
$x=\sqrt{|a|}$ is the vertical asymptote of the graph $Z(x,a)$, see
the example of the curve for $a=-1$ in box ${\rm f}$ of
Fig. \ref{electricfieldinnonamemodel}.  When $a$ tends to zero
remaining negative, the vertical asymptote shifts towards the line
$x=0$, and the solution $Z(x,a \to 0_{-})$ converts, finally, into the
Coulombian solution. At large values of $x$ the plots of the function
$Z(x,a<0)$ tend to the Coulombian curve $Z(x,0)$ for all values of the
quantity $a$.  Thus, for $a \leq 0$ the solutions $Z(x,a)$ are not
regular at all in the range $0 \leq x < \infty$.

\subsubsection{Metric functions}

To study the gravitational part of the solution, we observe that the
solution of (\ref{2eqdim}) for $N$ can be presented in quadratures
\begin{eqnarray}
N(x,a) = 1 -  \frac{1}{x} e^{a Z^2(x,a)} \left\{ K -
\right.\nonumber\\
\left.
\int_{\infty}^{x} d \xi e^{-a Z^2(\xi,a)} \left[  (\xi^2 + a) Z^2(\xi,a)
-  2 Z(\xi,a) {+} 2 a \xi Z(\xi,a) \frac{d}{d\xi} Z(\xi,a) \right]
\right\} \,,
\label{31eqdim}
\end{eqnarray}
where $Z(x,a)$ is supposed to be already
found. The constant of integration, $K$, can clearly
be related to the asymptotic mass of the object $M$. When $x \to
\infty$ and $Z(x \to \infty,a) \to \frac{1}{x^2}$, Eq.
(\ref{31eqdim}) yields that
$K =  \frac{r_M}{r_Q}$, as defined in (\ref{K0}).
On the other hand, searching for a solution $N(x,a)$, which is finite
at $x=0$, and taking into account that $Z(0,a)$ is finite, we should
require that
\begin{equation}
K = \int_{\infty}^{0} d \xi e^{-a Z^2(\xi,a)} \left[
(\xi^2 {+} a) Z^2(\xi,a)
{-}  2 Z(\xi,a) {+} 2 a \xi Z(\xi,a) \frac{d}{d\xi}
Z(\xi,a) \right]
\,. \label{KZreq}
\end{equation}
In this case the formula (\ref{31eqdim}) transforms into
\begin{eqnarray}
N(x,a) {=} \frac{1}{x} e^{a Z^2(x,a)}
\int_{0}^{x} d \xi e^{-a Z^2(\xi,a)}
\left[  (\xi^2 {+} a) Z^2(\xi,a)
{-}  2 Z(\xi,a) {+} 1 \right] \,,
\label{431eqdim}
\end{eqnarray}
providing
\begin{eqnarray}
N(0,a) = 1 + a Z^2(0,a) -  2 Z(0,a)  \,.
\label{531eqdim}
\end{eqnarray}
Since the electric field at the center satisfies the condition
\begin{eqnarray}
     1 - 2 a Z^2(0,a) + a^2 Z^3(0,a) = 0  \,,
\label{631eqdim}
\end{eqnarray}
(see (\ref{1eqdim})), then $N(0,a)$ can be rewritten as
$N(0,a)=1-\frac{1}{aZ(0,a)}$ in accordance with the first relation
from (\ref{regularity1}). Thus, for the family of solutions with
regular functions $N(r,a)$ and $E(r,a)$ the quantity $K$ is a
function of the value $Z(0,a)$, i.e., depends on the quantity $a$
according to the formula (\ref{KZreq}), $K=K(a)$.

Also, from (\ref{3eqdim}) for $\sigma$ one finds
\begin{equation}
\sigma(x,a) = \sigma(\infty,a) \exp\left\{-a \left[Z^2(x,a) -
Z^2(\infty,a) \right] \right\} \,, \label{21eqdim}
\end{equation}
with $Z(x,a)$ being found from (\ref{1eqdim}).  For the asymptotically
Coulombian branch we have to set $Z(\infty,a){=}0$ and thus
$\sigma(\infty,a)=1$.  Then, the function $1/A$ is taken from
$1/A=N$ and the function $B$ is taken from $A$ and $\sigma$,
$B=\sigma^2/A$.

\begin{figure}
\centerline{\includegraphics[width=6.76in,height=5.07in]{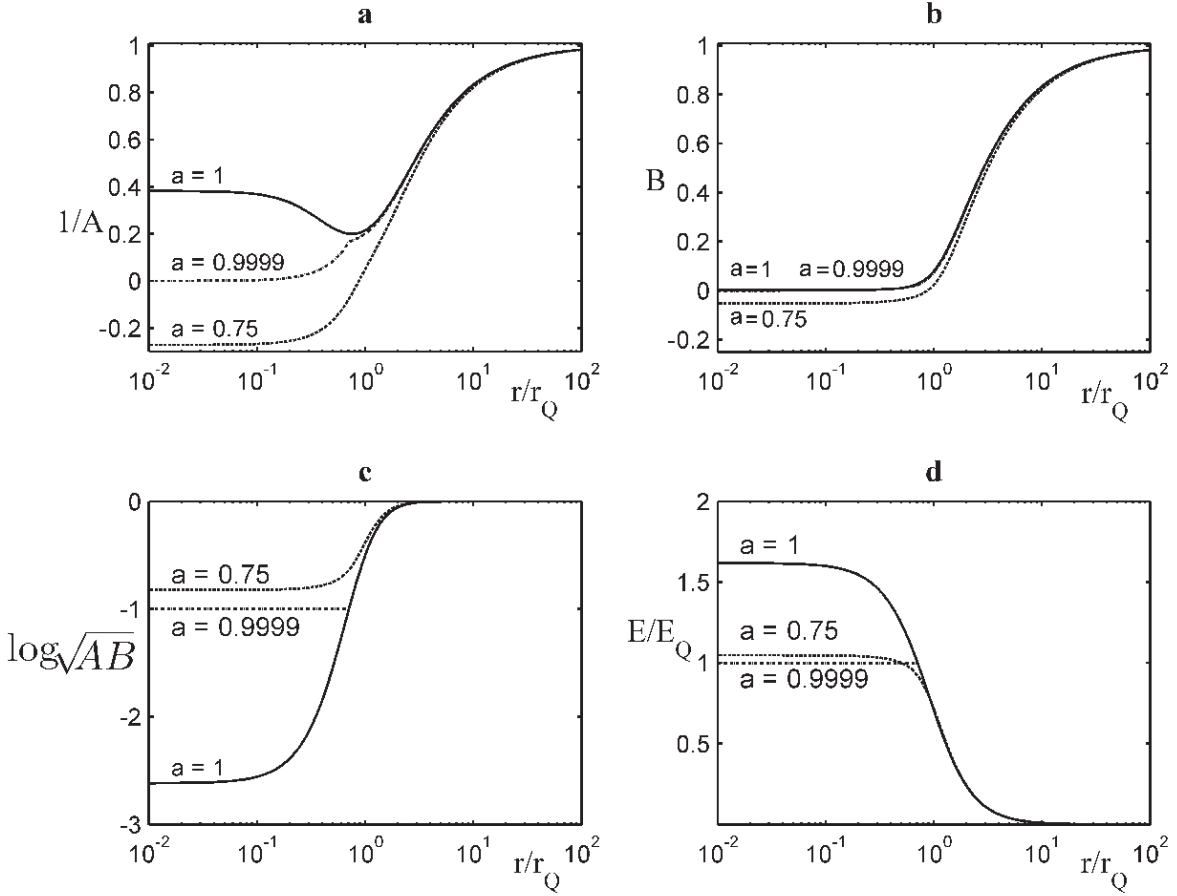}}
\caption{{\small Non-minimal solution of the
integrable model, with $q_1=-q$, $q_2=q$ and
$q_3=0$, of gravitational electrically charged objects  characterized by
$a=\frac{2q}{r^2_Q}$ --} {\small
Plots (a), (b), (c), and (d) depict the functions $\frac{1}{A}$, $B$,
$\log{\sqrt{AB}}$ and $E/E_{Q}$, respectively, as functions of
$x=\frac{r}{r_Q}$, for three typical values of the quantity $a$, when
$0 < a \leq 1$.  These functions are regular and take finite values at
$r=0$.  The solution with $a\equiv a_0 = 1$ is a solution without 
horizons,
since $\frac{1}{A}$ and $B$ are positive everywhere. Indeed the
$a\equiv a_0=1$ solution is a soliton of the model, the Fibonacci soliton. 
It
has a mild conical singularity at the center, and is
a solution without horizons.}}
\label{metricinnonnamemodel}
\end{figure}

In the Reissner-Nordstr\"om solution and in the previous discussed
model, i.e., the Gauss-Bonnet model (see Sections \ref{RNmodel} and
\ref{GBmodel}, respectively), we divided the solutions according to
$K>2$, $K=2$, and $K<2$.  Here it is no more convenient to make such a
division.  The reason is that $K$ is not a free quantity here, rather
$K=K(a)$. Since it is not a free quantity, we cannot classify the
models with respect to it, we can only calculate this quantity
(numerically) after solving the problem as a whole.  In the
Gauss-Bonnet model of section \ref{GBmodel} we could classify through
$K$ because the equations for $Z(x)$ and $N(x)$ cannot be decoupled,
thus, $Z(0,a)$ and $N(0,a)$ are connected. This means that we can
choose $K$ as a convenient quantity for the classification, the
dependent quantity being $Z(0,a)$.  Here, in this model, the solution
for $Z(x)$ satisfies a decoupled equation, the latter does not depend
on $N(x)$. Thus solving the decoupled equation for $Z(x,a)$, we can
classify the quantity $Z(0,a)$ as the independent one. Then, we obtain
$N(x,a)$ and, as we see, $N(0,a)$ depends on $E(0,a)$, an so $K$ is a
dependent quantity, $K=K(a)$.  An explicit example where $K(a)$ is
calculated is given below for the case $a\equiv a_0=1$.  So, as when
discussing the electric field, we again divide the analysis into three
cases, here with subcases.
\vskip 0.2cm
\noindent
{\it (i) $a>0$:} We should further divide into three subcases:

\noindent $ \infty>a>1$: The electric field is discontinuous or
irregular at the center. Since in this work we focus on regular
electric fields everywhere, although interesting, we do not discuss
these models here.

\noindent $a=1$, i.e, $a\equiv a_0=1$ :
We plot in Fig. \ref{metricinnonnamemodel}, boxes (a), (b), (c), (d),
the functions $1/A$, $B$, $\log\sqrt{AB}$, and $E/E_{Q}$,
respectively.  From the figure it is clear that $a{=}1$ is
characterized by the absence of horizons. For $a\equiv a_0=1$ the
solution is regular, or better, quasi-regular, and is a soliton of the
theory, the Fibonacci soliton. Due to its interest, the case $a\equiv
a_0=1$ will be solved next explicitly.

\noindent $1>a>0$: This case is interesting.  In this case we
plot in Fig. \ref{metricinnonnamemodel}, boxes (a), (b), (c), (d), the
functions $1/A$, $B$, $\log\sqrt{AB}$, and $E/E_{Q}$, respectively, for 
two
positive values of $a$ within this range, namely, $a=0.999, 0.750$.
From the figure it is clear that positive $a$ in this range gives one
horizon, and the solution possesses a quasi-regular center.  Since the
singularity at the center is a conical one, these black hole solutions
can be considered as quasi-regular solutions, and thus are of great
interest.  Note that extremal black holes have only one zero,
which in turn is a double zero. So from the figure above, the
non-minimal black holes do not characterize as extremal.
Rather, they are of the Schwarzschild type, with one horizon,
a spacelike singularity, but here, different from Schwarzschild,
the singularity is mild, it is a conical singularity.

\vskip 0.2cm
\noindent
{\it (ii) $a = 0$:} It is the Reissner-Nordstr\"om case.  For this case
the electric field is irregular at the center.

\vskip 0.2cm
\noindent
{\it (iii) $a < 0$:} For these cases the electric field is
discontinuous or irregular at the center, like the
Reissner-Nordstr\"om case, and we do not discuss
these models here.

\subsubsection{Exact solution: $a\equiv a_0=1$, the Fibonacci soliton}

Let $a\equiv a_0=1$. Then the cubic equation
(\ref{1eqdim}) takes the form
\begin{equation}
(Z-1) \left[(1 + x^2) Z^2 + (x^2 -1) Z - 1 \right] = 0 \,. \label{3to2}
\end{equation}
One sees that Eq. (\ref{3to2}) splits into one linear equation
and one quadratic equation. One branch of solutions of (\ref{3to2}),
the linear one, describes a constant electric field $Z_{\rm
const}(x,1)=1$, or, equivalently, $E(r) = E_Q$.  This branch is of no
great interest.  Another branch $Z_{\rm nonCoulomb}(x,1)$ is given by
the function $ Z_{\rm nonCoulomb}(x,1) =
\frac{1}{2(1+x^2)} \left[ 1-x^2 -\sqrt{x^4
+ 2x^2 +5} \right]$,  which is bounded.  The graph of
this function starts from $Z_{\rm nonCoulomb}(0,1) = -
\frac{\sqrt{5}-1}{2}$ and tends asymptotically to the line $Z=-1$. The
behavior of such electric field is not of Coulombian type, and will
be not discussed further.
Yet, there is a third branch.
The branch $Z_{\rm Coulomb}(x,1)\equiv Z(x,1)$ given by the formula
\begin{equation}
Z(x,1) = \frac{1}{2(1+x^2)} \left[ 1-x^2 + \sqrt{x^4
+ 2x^2 +5} \right] \,, \label{z3}
\end{equation}
describes a Coulombian type electric field. At $x \to \infty$, one has
$Z(x,1) \to \frac{1}{x^2}$, or equivalently, $E(r) \to
\frac{Q}{r^2}$. The graph of this function starts from $Z(0,1) =
\frac{\sqrt{5}+1}{2}$. Interesting to note that the starting points
$Z(0,1)$ are associated to the well-known Fibonacci series and the
``golden section'' $\phi \equiv \frac{\sqrt5 +1}{2} =
\frac{2}{\sqrt5-1} = 1.618...$.  For the Coulombian type solution, the
function $N(x,1)$, which we write simply as $N(x)$ when suitable, is
regular at the center only if the constant $K$ satisfies
(\ref{KZreq}). The corresponding quadrature for $N(x)$ is
\begin{equation}
N(x) = \frac{1}{2x \sigma(x)} \int_0^{x}
d \xi \ \sigma(\xi) \left[  \xi^2 + 3
- \sqrt{\xi^4 +2 \xi^2 +5} \right] \,,
\label{Nsi}
\end{equation}
where $\sigma(x)$ is given by (\ref{si}).
Clearly,  $N(\infty,1) =N(\infty) =1$ and $N(0,1)=N(0) =
\frac{3-\sqrt5}{2}$, so that $1-N(0) = \frac{1}{\phi} \equiv \phi-1$, and
the relations (\ref{regularity1}) are satisfied.
The plot of the function $N(x)$ for $a\equiv a_0=1$ is shown in  Fig.
\ref{metricinnonnamemodel}.
Clearly, the function $N(x)$ is positive in the interval $0 \leq x<
\infty$.
For the Coulombian type solution (\ref{z3}) the function
$\sigma(x,1)$, which we
write simply as $\sigma(x)$ when suitable, is
given by
\begin{equation}
\sigma(x) =  \exp\left\{- \ \frac{ 3  + (1-x^2) \sqrt{x^4
+ 2x^2 +5}+ x^4}{2 (1+x^2)^2} \right\} \,,
\label{si}
\end{equation}
with $\sigma(0,1)=\sigma(0)$ being equal to $\exp\{-(1+\phi) \}$.
Then one finds $1/A$ from $1/A=N$ and $B= \sigma^2/A$.
The function $B(x)$ is also positive,
and $B(0)= (1-\frac{1}{\phi}) e^{- 2(1+\phi)} \simeq 0.002032$.
Thus, this solution is a solution without horizons and is regular.
Although the curvature scalars diverge, the singularity
at the center is a mild one, it is a conical singularity.
The asymptotic mass $M$ of the object defined as
\begin{equation}
r_{M} \equiv 2GM = \lim_{r \to \infty} \left\{r \left[1 - B(r)
\right] \right\} \,,
\label{mass1}
\end{equation}
is represented in this case by the integral
\begin{equation}
M =  \frac{|Q|}{4\sqrt{G}} \int^{\infty}_0 d \xi \left[
\frac{1}{\sigma(\xi)}
- \xi \frac{\sigma^{\prime}(\xi)}{\sigma^2(\xi)} - \frac{1}{2}
\sigma(\xi)\left(\xi^2 + 3 - \sqrt{\xi^4 + 2 \xi^2 +5}
\right) \right]
\,.
\label{mass2}
\end{equation}
Numerical calculations give the value
\begin{equation}
M \simeq 0.442\,\frac{|Q|}{\sqrt{G}}\,,
\label{Mspecific}
\end{equation}
which yields in addition,
\begin{equation}
K= \frac{2 \sqrt{G} M}{|Q|} \simeq 0.884 \,.
\label{Kspecific}
\end{equation}
This is thus a very interesting solution. It is a soliton, in the
sense that is made of the very own fields of the theory, the
gravitational and electric fields, it is a solution without horizons,
and it is quasi-regular, with a conical singularity at the center.

Note also that the value $a\equiv a_0=1$ can be regarded
as a critical one. There
are two reasons for this. First, it is clear from
Fig. \ref{electricfieldinnonamemodel}, that for the unique case
$a\equiv a_0=1$
there exists a bifurcation point, in which the two branches of the
curve $Z(x,a)$, or $E(r,a)$, intersect.
When $a>1$, the Coulombian branch of the
electric field curve is discontinuous. When $ 0<a<1$, three continuous
regular branches exist. The second reason is that at $a\equiv a_0=1$ the
corresponding curve on Fig. \ref{metricinnonnamemodel} plays a
role of a separatrix; when $0<a<1$, desirable curves, continuous and
regular at the center, exist, otherwise they do not appear.

\subsection{Summary of the results}
\label{table}

\begin{figure}
[htmb]
\centerline{\includegraphics[width=6.75in,height=3.1in]{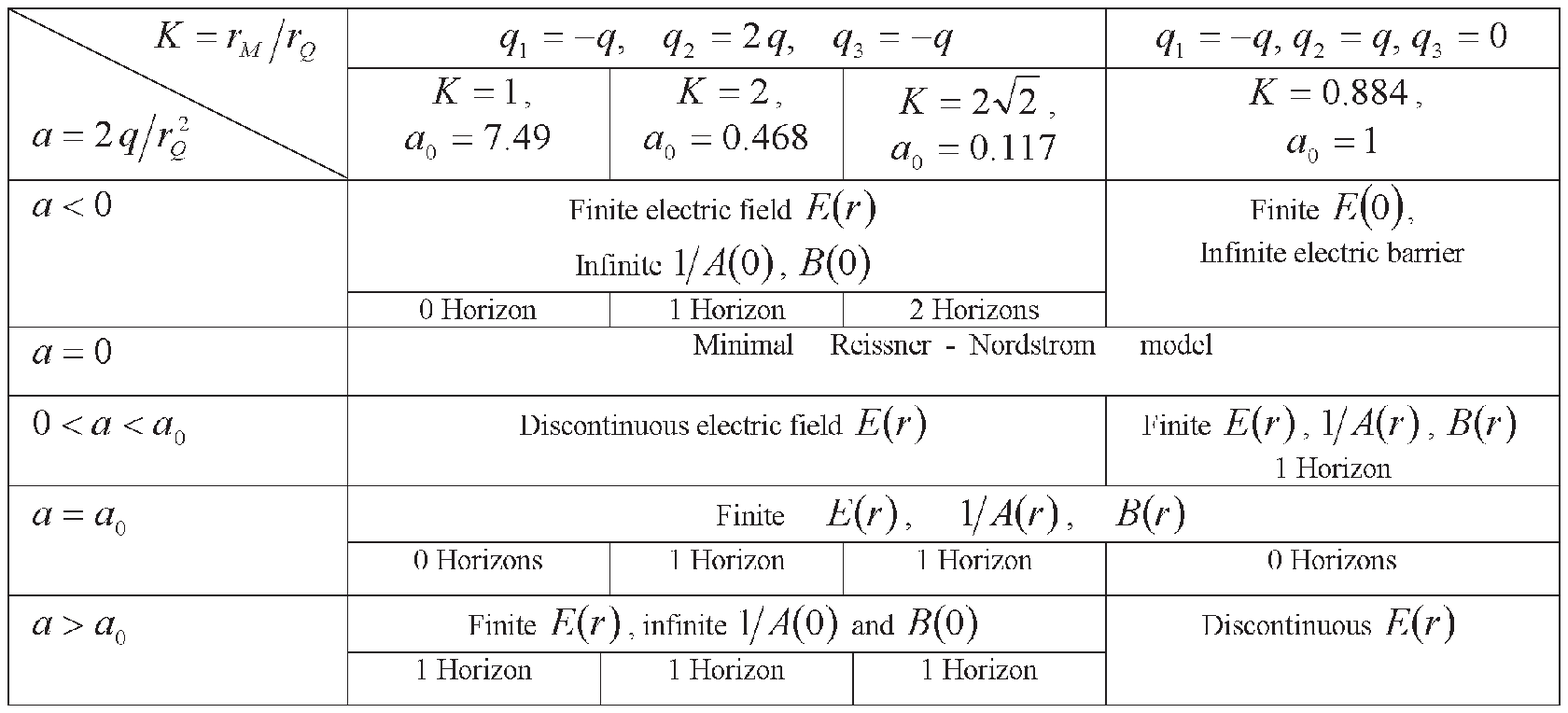}}
\caption{{\small In this table it is diplayed the main results
on the studied models.}}
\label{Table}
\end{figure}

In Fig. \ref{Table} we present a table in which the results of
the various models studied are summarized.

\section{Conclusions}
\label{conc}

We have shown that the original non-minimal Einstein-Maxwell theory
with three parameters $q_1$, $q_2$, and $q_3$, is reducible in natural
different ways to a theory with one parameter $q$ only, in which the
three parameters obey two relations between themselves. We have then
studied two special models for static spherically symmetric solutions
obeying the following requirements: the electric field $E(r)$ is
regular everywhere in the interval $0\leq r <\infty$, being Coulombian
far from the center.  From the solutions of this class we extract the
ones, for which the metric coefficients $\frac{1}{A(r)}$ and $B(r)$
are regular at the center $r=0$ and tend to one asymptotically as $r
\to \infty$.

The first non-minimal model, the Gauss-Bonnet model (with
$q_1\equiv -q$, $q_2=2q$, $q_3=-q$, $q$ free), displays charged black
hole solutions with one horizon only, when the dimensionless
non-minimal quantity $a$, with $a=2q/r_Q^2$
naturally appearing in the model, exceeds a
critical value $a_0$, $a>a_0$. Although the black hole is electrically
charged the solutions found have one horizon only, and are similar in
this connection to the Schwarzschild solution.  When $a<a_0$, the
solutions are discontinuous in the interval $0\leq r <\infty$, or
irregular at the center $r=0$.  Another main result in this model is that
there exists a unique solution, the solution for the critical value
$a{=}a_0$, which does not possess horizons and is characterized by
regular fields $E(r)$, $1/A(r)$ and $B(r)$, with $B(0){=}0$ and $A(0)
\neq 1$, although the curvature invariants blow at the origin.

The second model, the integrable model (with $q_1\equiv - q$, $q_2=q$,
$q_3=0$, $q$ free), is also characterized by one critical value
$a\equiv a_0=
1$ of the non-minimal quantity $a$. When $a<0$ or $a>1$, the
solutions are irregular. When $0<a<1$ one obtains black holes with
electric field regular everywhere and with only one horizon, like the
Schwarzschild solution. Finally, when $a\equiv a_0=1$, i.e.,
at the critical value
of the quantity $a$, there exists a solitonic solution with a conical
singularity at the center, but otherwise well behaved. This solution
can be called the Fibonacci soliton, since the well-known $\phi$
number ($\phi {=} \frac{\sqrt{5}{+}1}{2} \simeq 1.618$), associated
with the golden section, appears naturally in the expressions for the
central values of the electric field $G|Q|E(0) {=} \phi$ and the
metric coefficients are also related to $\phi$, namely
$1{-}\frac{1}{A(0)}{=}\frac{1}{\phi}$ and $B(0){=}
\left(1{-}\frac{1}{\phi} \right)\exp[{-}2(1{+}\phi)]$.

Summing up, we can say that the non-minimal curvature induced
interaction between the gravitational and electromagnetic fields provides
an electric field, of static spherically symmetric charged objects,
which is regular everywhere for different relations between the
coupling constants $q_1$, $q_2$ and $q_3$. As for additional
regularity of the metric coefficients $B(r)$ and $A(r)$, the
non-minimal interaction can provide models which have very specific,
critical, values for the coupling constants, in which the geometry has
at most a conical, and thus mild, singularity.  This is in line with
problem posed by Bardeen \cite{Bardeen}, where one should look for
theories with regular black hole solutions. We have partially solved
it within these models.

\section*{Acknowledgments}
AB thanks the hospitality of CENTRA/IST in Lisbon, and a special grant
from FCT to invited scientists.  This work was partially funded by
Funda\c c\~ao para a Ci\^encia e Tecnologia (FCT) - Portugal,
through project POCI/FP/63943/2005, and by Russian Foundation for Basic
Research through project RFBR 08-02-00325-a.

\end{document}